\documentclass[11pt,twocolumn,tight,times]{aastex631}

\usepackage{amsmath, mathrsfs}
\usepackage{multirow} 

\newcommand{\yr}{{\rm \,yr}}
\newcommand{\myr}{{\rm \,Myr}}

\newcommand{\mpc}{{\rm \,Mpc}}

\newcommand{\msun}{{M_\odot}}

\newcommand{\hz}{{\rm \,Hz}}

\newcommand{\kms}{{\rm \,km\,s^{-1}}}
\newcommand{\pyr}{{\rm \,yr^{-1}}}
\newcommand{\rmdex}{{\rm \,dex}}

\newcommand{\bulge}{_{\rm bulge}}

\newcommand{\obs}{_{\rm obs}}
\newcommand{\orb}{_{\rm orb}}
\newcommand{\ayr}{\mathcal{A}_{\rm yr}}
\newcommand{\gal}{_{\rm gal}}
\newcommand{\bbh}{_{\rm BBH}}
\newcommand{\gw}{_{\rm gw}}
\newcommand{\gwr}{_{\rm gw,r}}

\newcommand{\bh}{_{\rm BH}}
\newcommand{\bhpri}{_{\rm BH1}}
\newcommand{\bhsec}{_{\rm BH2}}
\newcommand{\galpri}{_{\rm gal1}}
\newcommand{\galsec}{_{\rm gal2}}
\newcommand{\calA}{{\mathcal{A}}}

\newcommand{\calK}{{\mathcal{K}}}

\newcommand{\calM}{{\mathcal{M}}}
\newcommand{\calN}{{\mathcal{N}}}
\newcommand{\calP}{{\mathcal{P}}}
\newcommand{\calR}{{\mathcal{R}}}

\newcommand{\rmcr}{_{{\rm c,r}}}
\newcommand{\rmcz}{_{{\rm c},z}}
\newcommand{\rmc}{_{\rm c}}
\newcommand{\rmr}{_{\rm r}}
\newcommand{\rmn}{_{\rm n}}
\newcommand{\psr}{_{\rm p}}

\newcommand{\tiomg}{{\tilde{\omega}}}
\newcommand{\tigam}{{\tilde{\gamma}}}
\newcommand{\tibet}{{\tilde{\beta}}}
\newcommand{\tialp}{{\tilde{\alpha}}}
\newcommand{\tieps}{{\tilde{\epsilon}}}

\newcommand{\aamaxd}{{\emph{delay-$\calA_1$}}}
\newcommand{\aamidd}{{\emph{delay-$\calA_{1/2}$}}}
\newcommand{\aamind}{{\emph{delay-$\calA_{1/4}$}}}
\newcommand{\aamaxn}{{\emph{nodel-$\calA_1$}}}
\newcommand{\aamidn}{{\emph{nodel-$\calA_{1/2}$}}}
\newcommand{\aaminn}{{\emph{nodel-$\calA_{1/4}$}}}
\newcommand{\bbmaxd}{{\emph{delay-$\calA^z_1$}}}
\newcommand{\bbmidd}{{\emph{delay-$\calA^z_{1/2}$}}}
\newcommand{\bbmind}{{\emph{delay-$\calA^z_{1/4}$}}}
\newcommand{\bbmaxn}{{\emph{nodel-$\calA^z_1$}}}
\newcommand{\bbmidn}{{\emph{nodel-$\calA^z_{1/2}$}}}
\newcommand{\bbminn}{{\emph{nodel-$\calA^z_{1/4}$}}}
\newcommand{\ccmaxd}{{\emph{delay-$\calA^{\rm e}_{\rm max}$}}}
\newcommand{\ccmidd}{{\emph{delay-$\calA^{\rm e}_{\rm med}$}}}
\newcommand{\ccmind}{{\emph{delay-$\calA^{\rm e}_{\rm min}$}}}
\newcommand{\ccmaxn}{{\emph{nodel-$\calA^{\rm e}_{\rm max}$}}}
\newcommand{\ccmidn}{{\emph{nodel-$\calA^{\rm e}_{\rm med}$}}}
\newcommand{\ccminn}{{\emph{nodel-$\calA^{\rm e}_{\rm min}$}}}

\newcommand{\klocsky}{{\calK_{\rm sky}^{\rm loc}}}
\newcommand{\klocsrc}{{\calK_{\rm src}^{\rm loc}}}
\newcommand{\kglbsky}{{\calK_{\rm sky}^{\rm glb}}}
\newcommand{\kglbsrc}{{\calK_{\rm src}^{\rm glb}}}
\newcommand{\nlocsky}{{\calN_{\rm sky}^{\rm loc}}}
\newcommand{\nlocsrc}{{\calN_{\rm src}^{\rm loc}}}
\newcommand{\nglbsky}{{\calN_{\rm sky}^{\rm glb}}}
\newcommand{\nglbsrc}{{\calN_{\rm src}^{\rm glb}}}
\newcommand{\nlocskymean}{{\left\langle\nlocsky\right\rangle}}
\newcommand{\nlocsrcmean}{{\left\langle\nlocsrc\right\rangle}}
\newcommand{\nglbskymean}{{\left\langle\nglbsky\right\rangle}}
\newcommand{\nglbsrcmean}{{\left\langle\nglbsrc\right\rangle}}

\begin{document}

\title{Pulsar timing array detections of supermassive binary black holes:
implications from the detected common process signal and beyond}
\shorttitle{GWB and BBH source detection via PTAs}
\shortauthors{Chen, Yu, \& Lu}

\author[0000-0001-5393-9853]{Yunfeng Chen} 
\affiliation{School of Astronomy and Space Science, University of Chinese
Academy of Sciences, Beijing 100049, China}
\affiliation{National Astronomical Observatories, Chinese Academy of Sciences,
Beijing, 100101, China; luyj@nao.cas.cn}
\affiliation{Kavli Institute for Astronomy and Astrophysics, and School of
Physics, Peking University, Beijing, 100871, China; yuqj@pku.edu.cn}

\author[0000-0002-1745-8064]{Qingjuan Yu}
\affiliation{Kavli Institute for Astronomy and Astrophysics, and School of
Physics, Peking University, Beijing, 100871, China; yuqj@pku.edu.cn}

\author[0000-0002-1310-4664]{Youjun Lu} 
\affiliation{National Astronomical Observatories, Chinese Academy of Sciences,
Beijing, 100101, China; luyj@nao.cas.cn}
\affiliation{School of Astronomy and Space Science, University of Chinese
Academy of Sciences, Beijing 100049, China}
\correspondingauthor{Qingjuan Yu}

\begin{abstract}
Pulsar timing arrays (PTAs) are anticipated to detect the stochastic
gravitational wave background (GWB) from supermassive binary black holes (BBHs)
as well as the gravitational waves from individual BBHs. Recently, a common
process signal was reported by several PTAs. In this paper, we investigate the
constraints on the BBH population model(s) by current PTA observations and
further study the detections of both the GWB and individual BBHs by
current/future PTAs. We find that the MBH--host galaxy scaling relation, an
important ingredient of the BBH population model, is required to either evolve
significantly with redshift or have a normalization $\sim0.86$--$1.1$\,dex
higher than the empirical ones, if the GWB is the same as the common process
signal. For both cases, the estimated detection probability for individual BBHs
is too small for a positive detection by current PTAs. By involving either the
constrained scaling relations or those empirical ones into the BBH population
models,  we estimate that the GWB may be detected with a signal-to-noise ratio
$\gtrsim3$ by the PTAs based on the Five hundred meter Aperture Spherical radio
Telescope (CPTA) and  the Square Kilometer Array (SKAPTA) after $\sim2-3$  (or
$\sim6-11$) years' observation,  if it is the same as (or  an order of magnitude
lower than) the common process signal. The detection time of individual BBHs by
CPTA and SKAPTA is close to that of the GWB detection.   We show that  the BBH
population model can be strongly constrained by the number and property
distributions of BBHs to be detected by future PTAs. 
\end{abstract}


\section{Introduction}
\label{sec:introduction}

Massive binary black holes (MBBHs, hereafter BBHs) are the natural products of
frequent galaxy mergers \citep[e.g.,][]{BBR80, Yu02} as a consequence of the
ubiquitous existence of massive black holes (MBHs) in the centers of galaxies
\citep[e.g.,][]{Ferrarese00, Gebhardt00, Tremaine02, KH13}. In the galaxy merger
remnants, the BBHs interact with surrounding stars and gas, which leads to
significant orbital decay 
(e.g., \citealt{BBR80, Yu02, Berczik06, Mayer07, Haiman09, Preto11, Khan11,
Khan13, Sesana13cqg}). 
Once the orbital separations of these BBHs become sufficiently small, e.g., at
mpc scales or smaller, they may emit large amounts of gravitational waves (GWs)
with frequencies at the nano-Hertz band ($\sim 10^{-9}-10^{-7}$\,Hz). Therefore,
the cosmic population of BBHs can be taken as one of the primary targets of the
pulsar timing arrays (PTAs). Two types of GWs from these BBHs are expected to be
detected by PTAs in the near future. One is the stochastic background (GWB)
produced by the incoherent combination of all the GW emissions from different
BBH systems, and the other is the continuous gravitational wave (CGW) emitted
from individual BBHs, which are resolvable if the signal stands out of the GWB.
Note that some cosmic relics from the early universe (e.g., see a recent
review by \citealt{Bian21}) or some more exotic sources (e.g., \citealt{WuYM22})
may also contribute to the nHz GWB, which are ignored in the present paper.

A number of PTAs have been operated for more than a decade, including the North
American Nanohertz Observatory for Gravitational Waves (NANOGrav,
\citealt{McLaughlin13}), the European PTA (EPTA, \citealt{Kramer13}), and the
Parkes PTA (PPTA, \citealt{Manchester13}). These three PTAs regularly update
their detection results, and combine together to form the  International Pulsar
Timing Array (IPTA, \citealt{Hobbs10, Verbiest16gwb}). Recently, a common
process signal has been reported by these PTAs \citep{Arzoumanian20cps,
Goncharov21cps, ChenSiyuan21cps, Antoniadis22cps}, which can be described by a
power-law spectrum with the power-law index close to that of the stochastic GWB.
However, its origin is still unclear as there has been no definitive evidence yet
supporting that the signal has a spatial correlation described by the
Hellings-Downs curve \citep{HD83}. Nevertheless, it may suggest that the
nano-Hertz GWB is close to be detected by PTAs. 

The GWB due to the cosmic population of BBHs, quantified by the characteristic
strain amplitude, $h\rmc$, has been theoretically estimated extensively in the
literature (e.g., \citealt{Phinney01, WL03, Sesana08, Sesana13, Ravi15,
Roebber16, Sesana16, Kelley17, CYL20bbh, Barausse20, Sykes22, Becsy22}). The GWB
strain amplitude estimated from different models can be different by more than
one order of magnitude, depending on the model assumptions and settings. This
suggests that those theoretical models can be effectively constrained if the GWB
signal is detected or a stringent upper limit on the GWB can be given by PTAs
(e.g., see \citealt{SB16, Middleton16, ChenSY17bbh, ChenSY17ecc, ChenSY19,
Middleton21, IV22, CC22}).

Many studies have also explored the detectability of individually resolvable CGW
sources, i.e., BBHs, in the literature (e.g., \citealt{Sesana09, Ravi15,
Mingarelli17, WangYan17, Kelley18, CYL20bbh, FengYi20}), though a consensus has
not been achieved. For example, based on a large ensemble of mock BBH
populations, \citet{Rosado15} concluded that the GWB is more likely to be
detected first. On the other hand, by coupling the cosmic population of galaxies
and MBHs from the Illustris cosmological hydrodynamic simulations to
semi-analytic models of binary mergers, \citet{Kelley18} concluded that
individual BBHs are at least as detectable as the stochastic GWB (see also
\citealt{Becsy22}). \citet{Mingarelli17} found that the detection probability of
local individual BBHs by the current generation of PTAs is negligible, e.g.,
$\ll 1\%$, by considering the possible BBH systems inside a sample of local
galaxies from observations.

As the GWB and the individual BBHs are expected to be detected by PTAs in the
near future,  many efforts have been made towards that goal in the past several
decades.  Indian PTA (InPTA, \citealt{Joshi18}) and China PTA (CPTA,
\citealt{LeeKJ16}) recently also joined the searching for nano-Hertz GWs, which
may be combined with IPTA together to improve the PTA sensitivity to the GWB and
individual BBHs. 
The MeerTime Pulsar Timing Array (MPTA, \citealt{Spiewak22MPTA}), which
will finish its initial five-year programme in July2024, has already played an
important role in the global efforts of IPTA in detecting GWs.
Even more powerful PTAs have been planning based on the Square Kilometer Arrays
(SKAPTA, \citealt{Smits09}) and the next generation Very Large Array (ngVLA,
\citealt{NANOGrav18}), which are anticipated to detect the GWB with high
signal-to-noise ratio (SNR) and detect many individual BBHs.  With these
detections, the cosmic formation and evolution of BBHs is expected to be further
constrained. However, how the BBH cosmic evolution model can be constrained
jointly by both the GWB and individual BBH detections has not been fully
explored.  

In this paper, we first investigate the possible constraints on the model of the
underlying BBH population that can be obtained by assuming that the recently
reported common process signal is indeed due to the GWB, and then explore the
detectability of individual BBHs based on the constrained model(s) and beyond.
Both the GWB strain amplitude and the occurrence rate of individual BBHs are
controlled by the cosmic distribution of BBHs, which is in turn controlled by
the cosmic BBH formation and evolution model. We adopt the cosmic BBH model in
\citet{CYL20bbh} (hereafter called CYL20), in which the following ingredients
are involved, i.e., the galaxy stellar mass function (GSMF), the merger rate per
galaxy (MRPG), the MBH--host galaxy scaling relation, and the time delays
between BBH coalescences and their host galaxy mergers.  For the constraints
given by PTA observations, we focus on the mass scaling relation between MBHs
and their host galaxies,  and leave other ingredients fixed.

This paper is organized as follows. In Section~\ref{sec:methods}, we briefly
introduce the main methods employed in this study. We present how to transfer
the possible GWB signal recently reported to the constraints on the model of the
BBH population in Section~\ref{sec:methods:constraints} and the method of
predicting the detection prospects of individual BBHs based on the constrained
BBH population model in Section~\ref{sec:methods:individuals}. In
Section~\ref{sec:misc}, we present the detailed model settings in this study,
such as the MBH-host galaxy properties, the PTA configurations and  their
sensitivities,  and the local sample of galaxies hosting the local BBH
population. The main results obtained in this study are given in
Section~\ref{sec:results}. Finally, the main conclusions are summarized in
Section~\ref{sec:conclusions}.

\section{Methods}
\label{sec:methods}

In Section~\ref{sec:methods:constraints},  we introduce our model for obtaining
constraints on the MBH--host galaxy scaling relation, by assuming that the
stochastic GWB has a strain amplitude the same as or a fraction of the common
process signal reported recently \citep{Arzoumanian20cps, Goncharov21cps,
ChenSiyuan21cps, Antoniadis22cps}. We explore both the case with the
redshift-independent scaling relations and that with the redshift-dependent
ones, highlighting the potential of using GWB ``observations'' (by using the
quota marks here we mean assumed observation of the GWB signal being the same as
or a fraction of the common process signal) to constrain the possible evolution
of the scaling relation with cosmic time. In
Section~\ref{sec:methods:individuals}, we introduce the method for predicting
the detection prospects of individual BBHs by current PTAs or future ones,
adopting both the MBH-host galaxy scaling relations constrained by the GWB
``observations'' and some empirical ones shown in the literature.

\subsection{Constraining the MBH-host galaxy relationship from the GWB}
\label{sec:methods:constraints}

The characteristic strain amplitude of the stochastic GWB in the PTA band,
$h\rmc$, produced by a cosmic population of BBHs at the GW frequency $f$ (in the
observer's rest frame) can be estimated as
\begin{eqnarray}
h\rmc^2(f)\simeq && \frac{4}{\pi} \frac{G}{c^2}f^{-2} 
\iiint dz dM\bh dq\bh \left|\frac{dt}{dz}\right| \nonumber\\
&& \times R\bh(M\bh,q\bh,z) \frac{1}{1+z} \left|\frac{dE}{d\ln f\rmr}\right|,
\label{eq:hc}
\end{eqnarray}
where $c$ is the speed of light, $G$ is the gravitational constant, 
$t$ is the cosmic time at redshift $z$ with $dt/dz$ being the
corresponding time derivative to $z$,
$f\rmr= (1+z)f$ is the frequency of the GW signal in the source's rest frame, 
and $|dE/d\ln f\rmr|$ is the GW energy per unit logarithmic rest-frame
frequency radiated by an inspiraling BBH with parameters $(M\bh, q\bh, f\rmr)$ 
(see Eqs.~30--34 in CYL20 and also the derivation of \citealt{Phinney01}). Note
that compared with Equation~(33) in CYL20,    
we set that the BBH evolution is at the gravitational radiation stage in Eq.~\eqref{eq:hc},
as we focus on the PTA band, where $f$ is greater than the turnover frequency of
the expected GWB shown in Fig.~19 in CYL20 and the coupling of the BBH orbital
evolution with surrounding environment is negligible.

In the above equation~(\ref{eq:hc}), $R\bh(M\bh,q\bh,z)$ represents the
coalescence rate of the BBHs, which is defined so that $R\bh(M\bh,q\bh,z) dt
dM\bh dq\bh$ represents the comoving number density of BBH coalescences occurred
during the cosmic time $t \rightarrow t+dt$, with the descendant mass of the two
component MBHs within the range $M\bh \rightarrow M\bh+dM\bh$ and the progenitor
mass ratio of the two MBHs within the range $q\bh \rightarrow q\bh+dq\bh$. The
characteristic strain amplitude $h\rmc (f)$ is determined by the BBH coalescence
rate, which is in turn determined by the cosmic BBH evolution model consisting
of several ingredients, including the GSMF $n\gal(M\gal,z)$, the MRPG
$\calR\gal(q\gal,z|M\gal)$, the MBH--host galaxy scaling relation, as well as
the dynamical evolution of BBHs within the galaxy merger remnants (see CYL20).
The GSMF is defined so that $n\gal(M\gal,z) dM\gal$ represents the comoving
number density of galaxies at redshift $z$ with stellar mass within the range
$M\gal \rightarrow M\gal+dM\gal$. The MRPG is defined so that
$\calR\gal(q\gal,z|M\gal) dt dq\gal$ represents the averaged number of galaxy
mergers with mass ratio in the range $q\gal \rightarrow q\gal+dq\gal$ within
cosmic time $t \rightarrow t+dt$ for a descendant galaxy with mass $M\gal$. With
these definitions, the BBH coalescence rate can be obtained through
\begin{eqnarray}
& R\bh (M\bh,q\bh,z(t)) 
= \frac{1}{N} \sum_{i=1}^{N} \iint dM\gal dq\gal \nonumber \\
& \times n\gal(M\gal,z_i) \calR\gal(q\gal,z_i|M\gal)  \nonumber \\
& \times p\bh(M\bh,q\bh|M\gal,q\gal,z_i)H(t-\tau_{a=0,i}),
\label{eq:Rbh}
\end{eqnarray}
where the BBH systems with coalescence timescales  $\tau_{a=0,i}$
($i=1,2,...,N$)  are generated by the Monte-Carlo method according to the
properties of the merged galaxies,  $H(t-\tau_{a=0,i})$ is a step function
defined in the way that $H(t-\tau_{a=0,i})=1$ if $t>\tau_{a=0,i}$ and
$H(t-\tau_{a=0,i})=0$ if $t\leq \tau_{a=0,i}$,  $z_i$ is the corresponding
redshift of the cosmic time $t-\tau_{a=0,i}$ (see derivations in Section~3.1 of
CYL20). In Equation~(\ref{eq:Rbh}),  $p\bh(M\bh,q\bh|M\gal,q\gal,z)$ is the term
through which the MBH--host galaxy scaling relation affects the BBH coalescence
rate. It is defined so that $p\bh(M\bh,q\bh|M\gal,q\gal,z) dM\bh dq\bh$
represents the probability of finding a BBH system with descendant total mass in
the range $M\bh \rightarrow M\bh+dM\bh$ and with progenitor mass ratio in the
range $q\bh \rightarrow q\bh+dq\bh$ if their host galaxies have descendant total
mass $M\gal $ and progenitor mass ratio $q\gal$ merged at redshift $z$.   To
obtain Equation~\eqref{eq:Rbh}, we substitute Equation~(15) in CYL20 into
Equation~(22) therein.  In the calculation, we ignore multiple galaxy major
mergers that could occur before the BBH coalescence since their host galaxy
merger, i.e., setting $P_{\rm intact}=1$ in Equation~(22) of CYL20, which is
plausible as the GWB at the PTA band is mainly contributed by galaxy or BBH
mergers within redshift lower than 2 (see Fig.~21 in CYL20).

We obtain  the probability $p\bh(M\bh,q\bh|M\gal,q\gal,z)$ in the following way.
Let $M\galpri$ and $M\galsec$ ($\leq M\galpri$) denote respectively the stellar
masses of the primary and secondary galaxies of a binary with total mass $M\gal$
and mass ratio $q\gal$. Similarly, let $M\bhpri$ and $M\bhsec$ ($\leq M\bhpri$)
denote respectively the masses of the primary and  secondary MBHs of a BBH
system with the total mass $M\bh=M\bhpri +M\bhsec $ (where the mass loss due to
the GW radiation is ignored) and mass ratio $q\bh$ ($\leq 1$). The probability
distribution function $p\bh$ can be calculated through
\begin{eqnarray}
& p\bh (M\bh,q\bh|M\gal,q\gal,z) = (M\bhpri^2/M\bh) \nonumber\\
& \times \left[p(M\bhpri|M\galpri,z) p(M\bhsec|M\galsec,z) \right.\nonumber\\
& \left.+ p(M\bhsec|M\galpri,z) p(M\bhpri|M\galsec,z) \right], 
\label{eq:pdf_BBH}
\end{eqnarray}
where $p(M\bh|M\gal,z)$ is defined so that $p(M\bh|M\gal,z) dM\bh$ represents the
probability of a galaxy with stellar mass $M\gal$ at redshift $z$ containing a
central MBH with mass within the range $M\bh\rightarrow M\bh+dM\bh$.

From Equation~(\ref{eq:pdf_BBH}), one observes that $p\bh$ in
Equation~(\ref{eq:Rbh}) is determined by the MBH--host galaxy scaling relation,
such as the $M\bh$--$M\bulge$ relation or the $M\bh$--$\sigma$ relation, where
$M\bulge$ and $\sigma$ represent the mass and stellar velocity dispersion of the
spheroidal components of the host galaxies (i.e., elliptical galaxies themselves
or bulges in spiral galaxies, throughout this work we use ``bulge'' to represent
both cases), respectively. Without loss of generality, we express the MBH--host
galaxy scaling relation as
\begin{equation}
\log M_{\rm BH,1} =
\tilde{\gamma}+\tilde{\omega}\log(1+z)+
\tilde{\beta}\log\sigma_{200}+
\tilde{\alpha}\log M_{\rm bulge,11},
\label{eq:scaling}
\end{equation}
with an intrinsic scatter of $\tieps$; 
and $M_{\rm BH,1}$, $\sigma_{200}$, and $M_{\rm bulge,11}$ represent the MBH
mass in unit of the solar mass $\msun$, the stellar velocity dispersion in unit
of $200\kms$, and the bulge mass in unit of $10^{11}\msun$, respectively. The
term $\tiomg\log(1+z)$ in the above equation describes the redshift evolution of
the scaling relation. If $\tiomg=0$, the scaling relation does not evolve  and
is independent of the redshift. Specifically, Equation~\eqref{eq:scaling}
reduces to the $M\bh$--$M\bulge$ relations if $\tibet=0$, $\tiomg=0$ and the
$M\bh$--$\sigma$ relation if $\tialp=0$, $\tiomg=0$. 

We leave $\tigam$ in Equation~\eqref{eq:scaling} as a free parameter to be
constrained, not only because the GWB strain amplitude is sensitive to the value
of $\tigam$, but also because the determination of $\tigam$ has not yet
converged in the literature. In addition, the derived relation with dynamical
mass measurements of MBHs may suffer significant selection bias
\citep{Shankar16}. The MBH--host galaxy scaling relation may evolve with
redshift as suggested by many authors (e.g., \citealt{McLure06, Salviander07,
Merloni10, ZLY12, Schulze14, Decarli18, DOnofrio21}), which may contain critical
information about the co-evolution of MBHs with their host galaxies. In this
work, we also demonstrate the potential of using GWB ``observations'' to set
strong constraints on the redshift evolution of the MBH--host galaxy
relationship.

The GWB strain amplitude can be evaluated according to the dynamical evolution
model for the cosmic BBHs as detailed in CYL20, which is mainly controlled by
the GSMF, MRPG, MBH--host galaxy scaling relation, and time delays between the
BBH coalescences and their host galaxy mergers.  Reversely,  we may apply the
Bayesian inference and the Markov Chain Monte Carlo (MCMC) method to extract
constraints on the model ingredients from the GWB ``observations'' in this
paper.  

The GWB strain amplitude can be well described by a power law in the PTA band
\citep{Phinney01}, except that  the small number variance becomes important at
the high-frequency end \citep{Sesana08} and  a bending emerges at the
low-frequency end due to the coupling of the BBH dynamical evolution by
interactions with its environment (CYL20) and/or highly eccentric BBH orbits
\citep[e.g.,][]{Enoki07}. Therefore, the observable can be represented by $\ayr$, which is
the characteristic strain amplitude at the observational frequency $1\pyr$,
i.e., $h\rmc=\ayr(f/1\pyr)^{-2/3}$. In this paper, we make the assumption that
the stochastic GWB produced by the cosmic population of BBHs has the same
amplitude as the common process signal recently reported by NANOGrav or smaller
than it by a factor of either $2$ or $4$, if not otherwise stated. When
constraining the BBH population model using $\ayr$, we focus on the MBH--host
galaxy scaling relation (cf.~Eq.~\ref{eq:scaling}), leaving the remaining
ingredients being fixed to fiducial choices. Details of the model settings are
described in Section~\ref{sec:misc:model}.

\subsection{Detections of individual BBHs}
\label{sec:methods:individuals}

In this subsection, we describe how to explore the detectability of individual
BBHs in the PTA band, based on the constraint given by the PTA observations on
the GWB or the empirical relations given by observations. We assume a BBH on the
circular orbit. In the source's rest frame, the BBH emits GWs at a frequency
twice the orbital frequency, i.e., $f\gwr=2f\orb$, which is then redshifted to
frequency $f\gw=(1+z)^{-1}f\gwr$ in the observer's rest frame, where $z$
represents the redshift of the BBH system. Note that $f$ and $f\gw$ denote the
GW frequencies of the stochastic GWB and the individual sources, respectively,
in this paper. The corresponding sky and polarization angular-averaged strain
amplitude of the BBH can be given by (see Eq. 55 in CYL20)
\begin{equation}
h_0 = \sqrt{\frac{32}{5}}\frac{1}{d_{\rm L}}
\left(\frac{G\calM\rmcz}{c^2}\right)^{5/3}
\left(\frac{\pi f\gw}{c}\right)^{2/3},
\label{eq:h0}
\end{equation}
where $d_{\rm L}$ is the luminosity distance of the BBH system, $\calM\rmcz=
(1+z)\calM\rmcr$ represents the redshifted chirp mass, and $\calM\rmcr=
M\bhpri^{3/5}M\bhsec^{3/5}/M\bh^{1/5}$ is the chirp mass of the BBH system in
the source rest frame.

We investigate the detectability of individual BBHs by PTAs for those BBHs in
the local universe and those as a cosmic population, respectively, by applying
the scaling relation between the MBH mass and host galaxy properties constrained
by the GWB ``observations'', if not otherwise stated, under the cosmic evolution
model of BBHs.
\begin{itemize}
\item
In the local universe (i.e., $z\simeq 0$), we have the observations of each
individual massive galaxy with well determined properties, e.g., stellar mass
and luminosity distance \citep{Arzoumanian21}. Some of them may even have well
determined MBH mass directly by gas/stellar kinematics. For those with direct
$M\bh$ measurements, we take it as the total mass of the BBH, if any. For those
without  direct $M\bh$ measurements, we set the BBH total mass the same as the
central MBH/BBH given by the MBH-host galaxy scaling relation constrained by the
GWB.  For each MBH in the local sample, it has a probability of being a BBH
system emitting GWs in the PTA band. Therefore, we can evaluate its conditional
probability distribution in the parameter space ($q\bh, f\gw$) at any given
$M\bh$, [i.e., $\calP\bbh(q\bh,f\gw,z=0|M\bh)$ in Eq.~\ref{eq:calPbbh} below;
see also CYL20]. A similar approach can be found in \citet{Mingarelli17}.
\item
For the high redshift universe, we do not have full information of all
individual galaxies. Therefore, we obtain the cosmic population of BBHs at high
redshifts based on the probability distribution of BBHs in the parameter space
$(M\bh,q\bh,f\gw,z)$ [i.e., $P\bbh(M\bh,q\bh,f\gw,z)$ in Eq.~\ref{eq:Pbbh}
below],  which can be obtained from the cosmic BBH formation and evolution model
in CYL20. In this way, one can get a comprehensive view of the individual BBH
detection by future PTAs, such as CPTA \citep{LeeKJ16} and SKAPTA
\citep{Smits09}, which are expected to detect a large number of BBHs. 
\end{itemize}
Below we denote the BBH systems derived from the two different approaches as the
``local population'' and the ``global population'', respectively, and describe
them in more details in Section~\ref{sec:methods:individuals:local} and
Section~\ref{sec:methods:individuals:global}, respectively.

\subsubsection{The local population of BBHs}
\label{sec:methods:individuals:local}

The probability distribution $\calP\bbh(q\bh,f\gw,z|M\bh)$ is defined in the way
that $\calP\bbh(q\bh,f\gw,z|M\bh) dq\bh df\gw$ represents the probability for a
BBH system with total mass $M\bh$ at redshift $z$ to have mass ratio in the
range $q\bh\rightarrow q\bh+dq\bh$ and emit GWs in the frequency range
$f\gw\rightarrow f\gw+df\gw$. We also define $\calR\bh(q\bh,z|M\bh)$ to be the
BBH specific coalescence rate (i.e., coalescence rate per BBH system) so that
$\calR\bh(q\bh,z|M\bh) dq\bh dt$ represents the averaged fraction of
coalescences that BBH systems, with total mass $M\bh$ and the mass ratio  in the
range $q\bh \rightarrow q\bh+dq\bh$, undergo during the cosmic time $t
\rightarrow t+dt$.  The above two quantities are related to each other through
\begin{eqnarray}
&&\calP\bbh (q\bh,f\gw,z|M\bh) = \nonumber\\
&&\calR\bh(q\bh,z|M\bh)\frac{1}{1+z}\left|\frac{dt\obs}{df\gw}\right|,
\label{eq:calPbbh}
\end{eqnarray}
where $dt\obs=(1+z)dt$ and
\begin{equation}
\frac{df\gw}{dt\obs} = \frac{96}{5}\pi^{8/3}
\left(\frac{G\calM\rmcz}{c^3}\right)^{5/3}f\gw^{11/3}.
\label{eq:dfgwdtobs}
\end{equation}
Note that an assumption of $\calR\bh$ being a constant during the period for the
BBH evolving from emitting GW at  $f\gw$ to its final coalescence is used to
obtain Equation~(\ref{eq:calPbbh}). In the observer's rest frame, the time to
coalescence can be given by
\begin{eqnarray}
\tau\obs
= && \frac{5}{256}\left(\frac{G\calM\rmcz}{c^3}\right)^{-5/3}
\left(\pi f\gw\right)^{-8/3} \nonumber\\
\simeq && 26\myr\left(\frac{8.7\times 10^8\msun}{\calM\rmcz}\right)^{5/3}
\left(\frac{10^{-9}\hz}{f\gw}\right)^{8/3},
\label{eq:tau_obs}
\end{eqnarray}
where the reference value for $\calM\rmcz$ is taken as $8.7\times 10^8\msun$,
the chirp mass of a binary composed of two equal-mass MBHs, each with a mass of
$10^9\msun$ at $z=0$. As clearly seen from this equation that $\tau_{\rm obs}$
is sufficiently small that the change of $\calR\bh$ over the coalescence time
can be safely ignored.

We define $n\bh(M\bh,z)$ to be the MBH mass function so that $n\bh(M\bh,z)
dM\bh$ represents the comoving number density of MBHs at redshift $z$ to have
mass in the range $M\bh\rightarrow M\bh+dM\bh$. Then the BBH specific
coalescence rate $\calR\bh$ and the BBH coalescence rate $R\bh$ are related to
each other through
\begin{equation}
\calR\bh(q\bh,z|M\bh) = \frac{R\bh(M\bh,q\bh,z)}{n\bh(M\bh,z)}.
\label{eq:calRbh}
\end{equation}
The mass function of MBHs is related to the mass function of their host galaxies
(i.e., the GSMF) through the MBH--host galaxy scaling relation, i.e., 
\begin{equation}
n\bh (M\bh,z) =
\int dM\gal n\gal(M\gal,z) p(M\bh|M\gal,z).
\label{eq:nbh}
\end{equation}

In short, given $p(M\bh|M\gal,z)$ constrained by the stochastic GWB,  we can
obtain the BBH coalescence rate $R\bh(M\bh,q\bh,z)$ (Eq.~\ref{eq:Rbh}) and the
MBH mass function $n\bh(M\bh,z)$ (Eq.~\ref{eq:nbh}), from which the specific
coalescence rate of BBHs $\calR\bh(q\bh,z|M\bh)$ can be further obtained through
Equation~(\ref{eq:calRbh}). Then,  we can give the conditional probability
distribution $\calP\bbh(q\bh,f\gw,z|M\bh)$ from the specific coalescence rate of
BBHs through Equation~(\ref{eq:calPbbh}).

\subsubsection{The global population of BBHs}
\label{sec:methods:individuals:global}

We now consider the global population of BBHs. Unlike the case of the local
population, in which the total mass of each BBH system $M\bh$ is fixed and thus
the conditional probability distribution $\calP\bbh(q\bh,f\gw,z|M\bh)$ is
needed, here instead we need to know the number distribution of BBHs in the
complete parameter space composed $(M\bh, q\bh, f\gw, z)$. We define this number
distribution as $P\bbh(M\bh,q\bh,f\gw,z)$ and it can be obtained through 
\begin{eqnarray}
&& P\bbh (M\bh,q\bh,f\gw,z) \nonumber\\
= && R\bh(M\bh,q\bh,z)\frac{1}{1+z} \left|\frac{dt\obs}{df\gw}\right|
\left|\frac{dV_{\rm c}}{dz}\right|,
\label{eq:Pbbh}
\end{eqnarray}
where $|dt\obs/df\gw|$ can be obtained from Equation~(\ref{eq:dfgwdtobs}),
$V_{\rm c}$ represents the comoving volume. Here $R\bh$ is also assumed to be a
constant during the period for the BBH evolving from emitting GW at $f_{\rm gw}$
to its final coalescence.

The global BBH population emitting CGW signals can be realized according to the
probability distribution $P\bbh(M\bh,q\bh,f\gw,z)$, which is different from the
local BBH population, for which the total mass of each system is given and
therefore the conditional probability distribution $\calP\bbh(q\bh,f\gw,z|M\bh)$
is employed to get the realizations.  Comparing Equations~\eqref{eq:calPbbh} and
\eqref{eq:Pbbh}, the two distributions differ from each other by the term
$n\bh(M\bh,z)$ (i.e., the MBH mass function) and the term $|dV_{\rm c}/dz|$.
Since the MBH mass function also depends on the scaling relation between MBH
mass and host galaxy properties, the realizations of the two populations may
change in different ways if the constraints on the relation change.

\section{Model settings}
\label{sec:misc}

In this section,  we introduce the model settings.  First, we briefly overview
the common process signal obtained by PTA observations and introduce our
settings on the PTA signals of the GWB in section~\ref{sec:PTAsignal}.  In
section~\ref{sec:misc:model}, we describe our settings on the galaxy properties
relevant to the cosmic evolution model for BBHs,  especially the MBH-host galaxy
scaling relation.  In section~\ref{sec:misc:sns}, we introduce our model setting
on PTA configurations and the estimates about their sensitivity curves.  We also
describe the local sample of MBHs in section~\ref{sec:misc:sample}.
 
\subsection{PTA signals}
\label{sec:PTAsignal}

The common process signal has been reported by NANOGrav
\citep{Arzoumanian20cps}, PPTA \citep{Goncharov21cps}, EPTA
\citep{ChenSiyuan21cps}, and IPTA \citep{Antoniadis22cps}. 
The reported median magnitudes of $\ayr$ by different PTAs range from ${\sim}2\times
10^{-15}$ to ${\sim}3\times 10^{-15}$, with a broad consistency at the
2--3$\sigma$ level \citep{ChenSiyuan21cps}.
Among the different values, we choose the one reported by NANOGrav as the
fiducial amplitude of the common process signal, which equals to
$1.92^{+0.75}_{-0.55}\times 10^{-15}$, with the quoted value being the median
and the quoted uncertainties for the $5\%$-$95\%$ confidence interval.
This is equivalent to a mean value of $-14.72$ and a standard deviation of
$0.09{\rm\, dex}$ for $\log_{10}\ayr$ if we assume that $\ayr$ follows a
log-normal distribution. Whether it represents the stochastic GWB is not clear,
as no conclusive evidence was obtained for  the measured spatial angular
correlation to be consistent with the Hellings-Downs curve \citep{HD83}. Even if
this signal is the real GWB, it is not clear whether this signal is all
contributed by the BBH population. 
For example, cosmological origin GWs, such as those induced by quantum
fluctuations at the inflation era, cosmic strings, domain walls, vacuum bubbles
(see  \citealt{Bian21}), may also contribute partly to the reported common
process signal.
Nevertheless, we assume the following three cases to consider the possible
constraints that may be obtained from the GWB ``observations'', i.e., the real
GWB or the fraction of the GWB contributed by the BBH population is assumed to
be 1) the same as, 2) a half of, or 3) a quarter of the common process signal.

\subsection{MBH and Galaxy Properties}
\label{sec:misc:model}

\begin{deluxetable*}{LCCCCCcl}
\tablecaption{Several empirical MBH--host galaxy scaling relations (see
Eq.~\ref{eq:scaling} and Section~\ref{sec:misc:model}) \label{tab:empirical}
adopted to study the detectability of individual BBHs.}
\tablehead{
\colhead{Model} & \dcolhead{\tigam} & \dcolhead{\tibet} & \dcolhead{\tialp} &
\dcolhead{\tieps} & \dcolhead{\log_{10}\ayr} & \colhead{Relation} &
\colhead{Reference}}
\startdata
\ccmaxd & 8.69 & \nodata & 1.17 & 0.29 & -15.28 & 
$M\bh$--$M\bulge$ & \citet{KH13} \\
\ccmidd & 8.23 & 3.96 & \nodata & 0.31 & -15.70 &
$M\bh$--$\sigma$ & \citet{Gultekin09} \\
\ccmind & 7.70 & 4.50 & 0.50 & 0.25 & -16.22 & 
$M\bh$--$\sigma$--$M\bulge$ & \citet{Shankar16}
\enddata
\end{deluxetable*}

The formation and evolution of the cosmic BBH population determines both the
stochastic GWB and individual CGW sources. As described in the introduction
section, the cosmic BBH formation and evolution model consists of several key
ingredients, including the GSMF, MRPG, MBH--host galaxy scaling relation, and
the time delays between BBH coalescences and their host galaxy mergers. The
first three can be obtained directly from observations though with some
uncertainties,  and the last one is  controlled by the BBH orbital evolution
\citep[e.g.,][]{Yu02}. It has been shown that the resulting GWBs can be
significantly different if adopting the MBH--host galaxy scaling relation given
by different authors (see CYL20). Therefore, we focus on the MBH--host galaxy
scaling relation (c.f., Eq.\,\ref{eq:scaling}),  and fix other related
ingredients, if considering the possible constraints on the cosmic BBH
population model from the GWB ``observations''. Specifically, we choose the GSMF
from \citet{Behroozi19} and the MRPG from \citet{Rodriguez-Gomez15}; to convert
the masses of the galaxies to the masses of their bulge components, we adopt the
prescription in \citet{Ravi15}; and we adopt that the galactic bulge have the
shape distribution described by \citet{Padilla08}. For the time delay,  we
employ the dynamical evolution model of BBHs developed in \citet{Yu02} to
calculate it as done in CYL20. For comparison, we also consider the extreme case
whithout time delay. 

We focus on the MBH--host galaxy scaling relation (cf.~Eq.~\ref{eq:scaling})
when considering the constraints that may be obtained from the GWB
``observations'' as mentioned above. In particular, we consider both the
redshift-independent and redshift-dependent $M\bh$--$M\bulge$ relation. For the
former case, we set $\tiomg=0$, $\tibet=0$, but keep $\tigam$, $\tialp$, and
$\tieps$ as free parameters to be constrained. For the latter case, we set
$\tigam=8.69$ (the same as that in the local $M\bh$--$M\bulge$ relation given by
\citealt{KH13}) and $\tibet=0$, but keep $\tiomg$, $\tialp$, and $\tieps$ as
free parameters to be constrained. We restrict the redshift evolution to be
within $z=3$, above which the scaling relation is assumed to be the same as that
at $z=3$. For the former case we assume that $\tigam$ has a flat prior
distribution within $[7.0,\,10.0]$, while for the latter case we assume that
$\tiomg$ has a flat prior distribution within $[-4.0,\,4.0]$. For both cases, we
assume that $\tialp$ and $\tieps$ have flat priors within $[0.8,\,1.2]$ and
$[0.0,\,0.6]$, respectively. 
We note here that in principle, one could set all the five parameters
(i.e., $\tigam$, $\tiomg$,  $\tibet$, $\tialp$, and $\tieps$) as free ones, and
obtain constraints on them by the GWB spectrum observations.

Furthermore, we also consider those cases that the MBH--host galaxy scaling
relation is the same as that directly determined by the MBHs in nearby galaxies
with dynamical mass measurements, without considering the constraints from the
common process signal detected by PTAs. In those cases, we also assume the
scaling relation does not evolve with redshift. Table~\ref{tab:empirical} lists
several empirical relations, which are  adopted to lead to the maximum, median,
and minimum GWB strain amplitudes estimated in CYL20, respectively. Note that
these empirical relations are only adopted as part of the cases to predict the
detectability of individual sources.

As described in Sections~\ref{sec:PTAsignal} and \ref{sec:misc:model},  various
models are considered in this study.  For clarity, we describe the notations of
the models in the following way (see
Tabs.~\ref{tab:constraints}--\ref{tab:tng}). Regarding the scaling relation,
these models can be divided into three groups as labeled by $\calA$, $\calA^z$,
and $\calA^{\rm e}$, respectively, corresponding to those admitting the
$z$-independent and $z$-dependent scaling relations constrained by the
stochastic GWB,  and those admitting the scaling relations empirically
determined. For those models admitting the constrained scaling relations,
$\calA_1$ , $\calA_{1/2}$ , and $\calA_{1/4}$ (or $\calA^z_1$, $\calA^z_{1/2}$,
and $\calA^z_{1/4}$) correspond to the cases in which the stochastic GWB is
assumed to have the amplitude equal to the common process signal scaled by a
factor of $1$, $1/2$, and $1/4$, respectively. For those models admitting the
empirical scaling relations, \emph{$\calA^{\rm e}_{\rm max}$}, \emph{$\calA^{\rm
e}_{\rm med}$}, and \emph{$\calA^{\rm e}_{\rm min}$} represent the model that
produces the maximum, medium, and minimum amplitudes of the stochastic GWB
listed in CYL20, corresponding to the $M\bh$--$M\bulge$ relation from
\citet{KH13}, the $M\bh$--$\sigma$ relation from \citet{Gultekin09}, and the
$M\bh$--$\sigma$--$M\bulge$ relation from \citet{Shankar16}, respectively.
Regarding the time delays between BBH coalescences and their host galaxy
mergers, we use \emph{``delay''} and \emph{``nodel''} to indicate those cases in
which the time delays are included and ignored, respectively. 
Note that the time delays given by CYL20 (Fig. 8 therein) have a wide
distribution from $\sim 10^8$\,yr to $10^{11}$\,yr with a peak around a few Gyr
depending on the BBH total mass and mass ratio.

\subsection{PTAs}
\label{sec:misc:sns}

\begin{deluxetable}{lCCCC}
\tablecaption{Assumed parameters for some PTAs based on FAST and SKA (see Section~\ref{sec:misc:sns}).  \label{tab:PTAs}}
\tablehead{
\multirow{2}{*}{
{Name} } &  
\dcolhead{\calN\psr} & 
\dcolhead{\sigma_a} & \dcolhead{T} & \dcolhead{\Delta t}\\
& &  \dcolhead{\rm (ns)} & \dcolhead{\rm (yr)} & \dcolhead{\rm (yr)}}
\startdata
conservative-CPTA      &   50 & 100 &  5 & 0.04 \\
conservative-SKAPTA &  100 & 100 &  5 & 0.04 \\
optimistic-CPTA           &  100 &  20 & 20 & 0.02 \\
optimistic-SKAPTA      & 1000 &  20 & 20 & 0.02 \\
\enddata
\end{deluxetable}

\begin{figure}[!htb]
\centering
\includegraphics[width=0.45\textwidth]{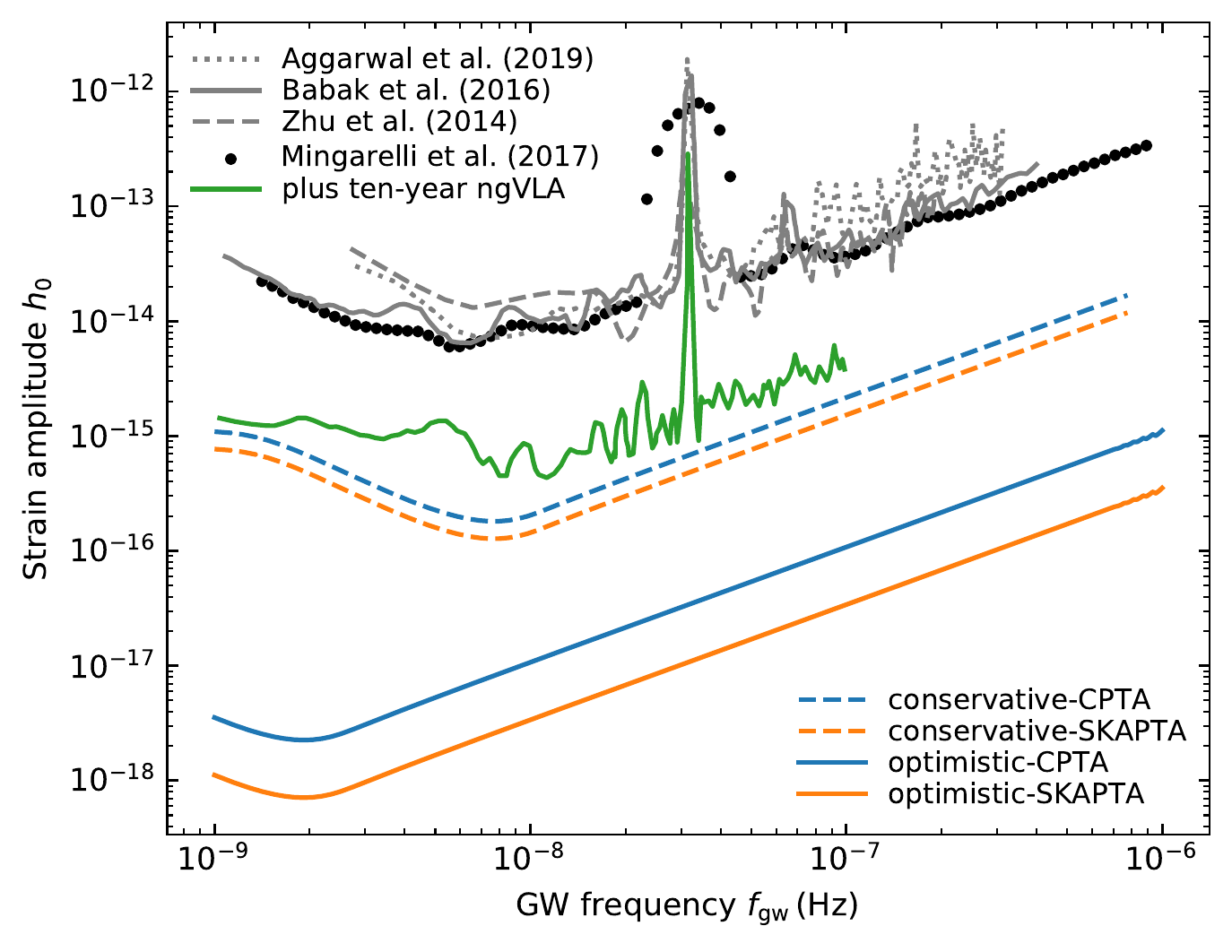}
\caption{The expected sensitivity curves for individual BBH detection by some
(future) PTAs (in Tab.~\ref{tab:PTAs}).  The blue (orange) dashed and solid
curves show the expected sensitive curves (with a threshold SNR $\rho_{\rm
th}=1$) for the conservative-CPTA and optimistic-CPTA (conservative-SKAPTA and
optimistic-SKAPTA),  respectively. The other curves are the reference curves.
The grey dotted,  solid,  and dashed curves show the sky-averaged $95\%$ upper
limits on the strain amplitude of individual BBHs based on the current PTA
observations,  i.e.,  based on NANOGrav 11-year timing data
\citep{Aggarwal19cgw},  EPTA DR1 timing data \citep{Babak16cgw}, and PPTA DR1
timing data \citep{ZhuXingjiang14cgw}, respectively.  The black dots also show
the sky-averaged 95\% upper limits based on EPTA DR1 timing data
\citep{Desvignes16} given by \citet{Mingarelli17}. The green curve shows the
simulated sky-averaged $95\%$ upper limits for the current NANOGrav program
followed by $10$ years of ngVLA observations \citep{NANOGrav18}.   See
Section~\ref{sec:misc:sns} for details.} 
\label{fig:sns4cgw_demo}
\end{figure}

\subsubsection{PTA configurations}
\label{sec:PTAconfig}

Here we introduce the configurations for those PTAs considered in this paper.
We consider both the PTAs that have been operated for many years (e.g., EPTA,
NANOGrav, and PPTA) and those new/future ones using FAST/SKA to explore the
detection prospects of individual BBHs. For the former PTAs, the upper limits on
individual BBH detection can be obtained by using the available data. For the
latter ones, we consider four PTA configurations by using FAST and SKA, with
properties (i.e., number of millisecond stable pulsars $\calN_{\rm p}$, timing
precision $\sigma_a$, cadence $1/\Delta t$, and observational time span $T$)
listed in Table~\ref{tab:PTAs}. We adopt two types of configurations for both
CPTA and SKAPTA: one is a conservative configuration and the other is an
optimistic configuration.   Note that \citet{Porayko18} listed the expected
white noise of the ten best PPTA pulsars observed in the FAST/SKA era in their
Table~4,  where all the ten pulsars have $\sigma_a$ well below $100\,{\rm ns}$
with a mean value of ${\sim} 30\,{\rm ns}$ and four pulsars have $\sigma_a$ even
below $20\,{\rm ns}$. Therefore, we assume that the conservative configurations
have a timing precision of $\sigma_a =100\,{\rm ns}$ and the optimistic
configurations have $\sigma_a=20$\,ns.   

In this study,  we also investigate the evolution of the detection prospects of
both types of GW signals with the PTA observation time $T$, as presented in
Section~\ref{sec:results:individuals:tng}. For these investigations, we still
adopt the PTA configurations listed in Table~\ref{tab:PTAs} except that the PTA
observation time is set as a free parameter (not fixed to those values listed in
Tab.~\ref{tab:PTAs}).  We append a star symbol to the PTA names (e.g.,
conservative-CPTA$^\ast$) to specially denote these cases.

\subsubsection{PTA sensitivity curves}
\label{sec:PTAsen}

The sensitivity curves for these PTAs can be estimated based on the
cross-correlation method (see Eq.~86 in \citealt{GLY22nf}, and see also
\citealt{Moore15pta}), i.e.,
\begin{equation}
\rho^2 = \calN\psr(\calN\psr-1)\frac{\chi^4 h_0^4(f)T^2}{S\rmn^2(f)}.
\label{eq:snr4cgw}
\end{equation}
Here $\rho$ denotes the SNR, $\calN\psr$ is the number of pulsars used by the
PTA, $\chi$ is a geometric factor and it is $1/\sqrt{3}$ obtained by the
far-field approximation for the distant sources. In Equation~\eqref{eq:snr4cgw},
$S\rmn(f)= 8\pi^2 f^2 \sigma_a^2 \Delta t$ is the noise power spectrum density
(PSD) and it is related to the noise amplitude $h\rmn(f)$ through
\begin{equation}
h\rmn(f) = \sqrt{fS\rmn(f)+h\rmc^2(f)},
\label{eq:hnf}
\end{equation}
where the stochastic GWB is regarded as a source of noise. If the GWB strain is
well determined, then the term $h\rmc^2(f)$ in Equation~(\ref{eq:hnf}) may be
dropped and thus $h\rmn(f) \simeq \sqrt{fS\rmn(f)}$. 
Note that when PTAs (such as the optimistic configurations of CPTA or
SKAPTA in Section~\ref{sec:results:individuals:tng}) are so sensitive that many
individual sources, crowding in each resolvable frequency bin, have
characteristic strain higher than the sensitivity curves, new challenges arise on
how to extract the signal of each individual BBH and how many individual BBHs
can be identified in each frequency bin. This is beyond the scope of the current
work, but of interest for future studies. With
Equation~(\ref{eq:snr4cgw}), we can estimate the sensitivity curve for a given
PTA configuration through
\begin{equation}
h_{0,{\rm th}}(f) = \left[\frac{\rho_{\rm th}^2}
{\calN\psr(\calN\psr-1) \chi^4}\right]^{\frac{1}{4}}
\frac{h\rmn(f)}{\sqrt{fT}}.
\label{eq:sns4cgw}
\end{equation}
According to Equation~(\ref{eq:sns4cgw}), we can see that $h_{0,\rm th} \propto
\sigma_a$, $h_{0,\rm th} \propto T^{-1/2}$ and $h_{0,\rm th} \propto \Delta
t^{1/2}$. When the number of pulsars is large, we also have $h_{0,\rm th}
\propto \calN\psr^{-1/2}$ approximately.  The sensitivity of a PTA can be
improved by improving the timing precision,  increasing the number of pulsars,
the cadence,  and the observation time. 

For the parameterized PTA configurations listed in Table~\ref{tab:PTAs}, we
apply  $\rho_{\rm th}=1$ in Equation~\eqref{eq:sns4cgw} to evaluate their
sensitivity curves and show the results in Figure~\ref{fig:sns4cgw_demo}. We
define a BBH as ``detectable'' if its SNR is above $3$. Below when studying the
detection statistics of individual BBHs, we evaluate both the \emph{detection
probability} and (average) \emph{detection number} of these sources. The former
describes the fraction of GW sky realizations containing at least one detectable
source, while the latter describes the average number of detectable sources in
each realization. Assuming that the occurrence rate of individual BBHs follows
the Poisson distribution when the number of ``detectable'' sources is small, a
detection probability of $95\%$ corresponds to a detection number of ${\sim} 3$,
and a detection number of $1$ corresponds to a detection probability of
${\sim}63\%$.

Note that our treatments to these parameterized PTA configurations as described
above are different from those with detection upper limits given by current
PTAs, e.g., the EPTA sensitivity skymap taken from \citet{Mingarelli17}. For the
latter ones, we use the $95\%$ upper limits as their sensitivity curves.  For
those BBHs with strain amplitudes $h_0$ above the sensitivity curves in this
case, we regard them as detectable sources. The \emph{detection probability} and
\emph{detection number} are the same as the parameterized PTA configuration
case.

Figure~\ref{fig:sns4cgw_demo} shows the expected sensitivity curves of those
PTAs listed in Table~\ref{tab:PTAs}. This figure also plots the simulated
sky-averaged $95\%$ upper limits for the current NANOGrav program followed by 10
years of ngVLA observations \citep{NANOGrav18}. Note that the curves for the
current PTAs and ngVLA denote the $95\%$ upper limits, while the curves for
those PTAs listed in Table~\ref{tab:PTAs} denote the PTA sensitivity on $h_0$
with a threshold SNR $\rho_{\rm th}=1$ (see Eq.~\ref{eq:sns4cgw}). As expected,
future PTAs may achieve a sensitivity $1-2$ orders of magnitude higher than the
current ones. One may pay attention to the upper limits constrained by those
currently available PTA observations. For example, the 95\% upper limit skymap
on the strain amplitude of individual BBHs was obtained by \citet{Mingarelli17}
at $86$ frequencies log-uniformly distributed between $10^{-9}\hz$ and
$10^{-6}\hz$ (the dots shown in Figure~\ref{fig:sns4cgw_demo}), based on the
EPTA DR1 data set \citep{Desvignes16}. Note that we exclude the bad pixels and
some pixels with suspicious upper limit values in the skymap, i.e., those with
negative upper limit values and those with declination satisfying $|\delta|>
75^\circ$ and with values below $1\times 10^{-15}$ when $|\delta|< 55^\circ$ and
below $4\times 10^{-15}$ when $|\delta|> 55^\circ$.

\subsubsection{SNR of the GWB expected by PTAs }

We also consider the expected SNR of the GWB for any given PTA, which can be
estimated as (see Eq.~23.69 in \citealt{Maggiore18}; \citealt{Siemens13})
\begin{equation}
{\rm SNR} = \left[\sum_{ab}\int_{f_l}^{f_h} df\frac{2T \zeta^2_{ab}
P^2_g(f)}{[2\sigma_a^2\Delta t+P_g(f)][2\sigma_b^2\Delta t+P_g(f)]}
\right]^{1/2},
\label{eq:snr4gwb}
\end{equation}
where $f_l=1/T$, $f_h=1/\Delta t$, $P_g(f) = h\rmc^2(f)/(12\pi^2f^3)$ (see
Eq.~23.51 in \citealt{Maggiore18}), $\sigma_a$ and $\sigma_b$ are the timing
precision of pulsars $a$ and $b$, respectively, $\zeta_{ab}=\frac{3}{2} C_{ab}$
with $C_{ab}$ being the angular correlation function, i.e., the Hellings-Downs
curve \citep{HD83}. Note that the value of $\zeta_{ab}$ depends on the relative
angle between the two pulsars $\theta_{ab}$ (see Eq.~23.48 in
\citealt{Maggiore18}). If PTA pulsars follow an isotropic distribution in the
sky, $\theta_{ab}$ follows a probability distribution with
$dP(\theta_{ab})/d\theta_{ab} = \frac{1}{2}\sin \theta_{ab}$. Therefore, when
evaluating the expected SNR of the GWB, we replace $\sum_{ab} \zeta_{ab}^2$ in
Equation~(\ref{eq:snr4gwb}) with
\begin{equation}
\left\langle\sum_{ab}\zeta_{ab}^2\right\rangle = 
\frac{\calN_{\rm p}(\calN_{\rm p}-1)}{4} 
\int_0^{\pi} \zeta_{ab}^2   \sin \theta_{ab} d\theta_{ab},
\label{eq:zeta2}
\end{equation}
under the assumption $\sigma_a\simeq \sigma_b$. 
Note that there are more pulsars in the Galactic plane and thus the real
distribution of PTA pulsars may deviate from the isotropic one. We have checked
and found that this may only lead to minor effects on our results.

\subsection{The local sample of MBHs}
\label{sec:misc:sample}

The constrained BBH population model can be applied to a local sample of MBHs
for exploring the detection statistics of individual BBHs in the local universe
as described in Section~\ref{sec:methods:individuals:local}. To do this, we
adopt the sample of nearby galaxies as compiled by \citet{Arzoumanian21}. Below
we briefly describe the sample.  

The local galaxy sample are selected by using the 2MASS $K$-band ($2.2\,\mu m$)
with apparent magnitude
$m_K\leq 11.75$, with a completeness of $97.6\%$ to a distance of $300\mpc$. It
consists of $43,533$ galaxies with spectroscopic redshifts in the 2MASS Redshift
Survey (2MRS, \citealt{Huchra12}), and represents a census of galaxies in the
local universe \citep{Arzoumanian21}. This sample has expanded in both the
volume and the sample size considerably, compared with the local galaxy sample
adopted in \citet{Mingarelli17}.

In the local sample, about $20\%$ of the galaxies ($8,625$) have directly
measured distances according to the Cosmicflows-3 catalog \citep{Tully16}. The
distances in the catalog are measured with high-quality methods, such as the
standard candles using the Cepheids, the tips of the red giant branch, or the
type Ia supernovae, and also the methods relying on some empirical relations
(e.g., the Tully-Fisher relation of spiral galaxies, the fundamental plane of
elliptical galaxies, and the surface brightness fluctuations). In addition, the
distances of $\sim$10\% of the galaxies ($4,533$) can be obtained from the
galaxy group catalog of \citet{Crook07}. For the remaining galaxies,  their
distances are estimated based on their redshifts. Within the local volume,
corrections need to be made to the spectroscopically measured velocities when
applying the Hubble's law to estimate distances, since they are significantly
affected by the peculiar motions of the galaxies. The prescription of
\citet{Mould00} is adopted to make such corrections.

We assign a mass to the central MBH (BBH) of each galaxy in the local sample.
Among these local galaxies, $77$ have direct dynamical mass measurements and
$29$ have mass measurements via the reverberation mapping method. For these
MBHs, their masses are fixed to the measured values, regardless of the changes
of the constraints on the MBH--host galaxy scaling relations
(Eq.~\ref{eq:scaling}). For the remaining MBHs, the MBH--host galaxy scaling
relations are applied to evaluate their masses, such as the $M\bh$--$\sigma$
relation ($2,206$) and the $M\bh$--$M\bulge$ relation ($41,221$) as quoted from
\citet{MM13} (see Tab.~2 of \citealt{Arzoumanian21} for details). Note that for
those MBHs without direct mass measurements, we adopt either the MBH--host
galaxy scaling relations constrained by the GWB ``observations'' or the
empirical ones given by the local MBH and galaxy observations to estimate the
MBH masses. When necessary, we adopt the $M\bulge$--$\sigma$ relation of
\citet{Gallazzi06} to make the conversion between the two quantities.

\section{Results}
\label{sec:results}

We present our main results in this section, which are divided into two parts.
As the GWB is expected to be detected before individual sources, in the first
part, we focus on the constraints that can be extracted from the stochastic GWB
``observations'', assuming that it is either the same as or a fraction of the
common process signal recently detected by several PTAs
\citep[][]{Arzoumanian20cps, Goncharov21cps, ChenSiyuan21cps, Antoniadis22cps}.
In the second part, we focus on applying the new constraints on the MBH--host
galaxy scaling relations obtained in the first part,  as well as the empirical
ones,  given by the local MBH and galaxy observations to explore the detection
prospects of individual BBHs by current and future PTAs. When the individual
sources were detected after the detection of the GWB, the BBH population model
can be further constrained by using both detections. 

\subsection{Possible Constraints on the MBH--host galaxy scaling relation by the GWB ``observations''} 
\label{sec:results:constraints}

We adopt the MCMC method ({emcee}; \citealt{emcee}) 
to obtain the constraints on the MBH--host galaxy scaling relation by matching
the GWB produced from each BBH evolution model to the ``observational'' ones
(assuming from BBHs), which is assumed to be either the same as or a fraction of
the common process signal.  We first consider the cases that the MBH-host galaxy
scaling relation is redshift-independent and redshift-dependent in
Sections~\ref{sec:zindep} and \ref{sec:zdepend}, respectively. The GWB
contributed by BBHs is assumed to be the same as the common process signal, 
of which the likelihood adopted is a log-normal distribution of the
NANOGrav posteriors. We obtain the constraints through $\ayr$ by assuming the
GWB spectrum follows the canonical $f^{-2/3}$ power-law. 
In Section~\ref{sec:DiffA},  we also further consider the cases that the GWB
contributed by BBHs is a fraction ($1/2$,  or $1/4$) of the common process
signal for the models with both redshift-independent and redshift-dependent
MBH-host galaxy scaling relations. 
In Section~\ref{sec:compare}, we compare our results of constraining the
cosmic BBH population using PTA observations with some literature works.

\begin{figure*}[!htb]
\gridline{
\fig{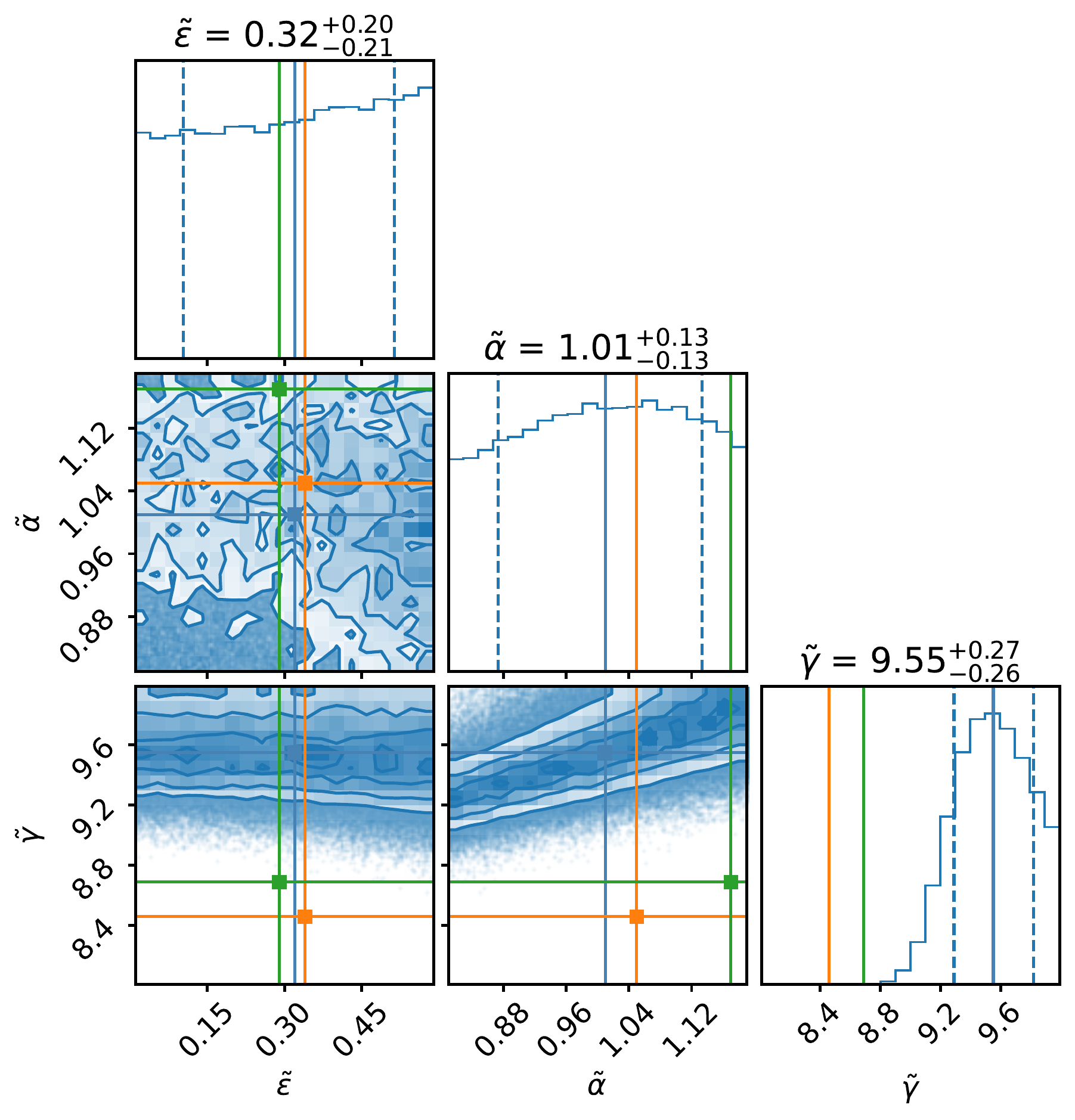}{0.45\textwidth}{(a)}
\fig{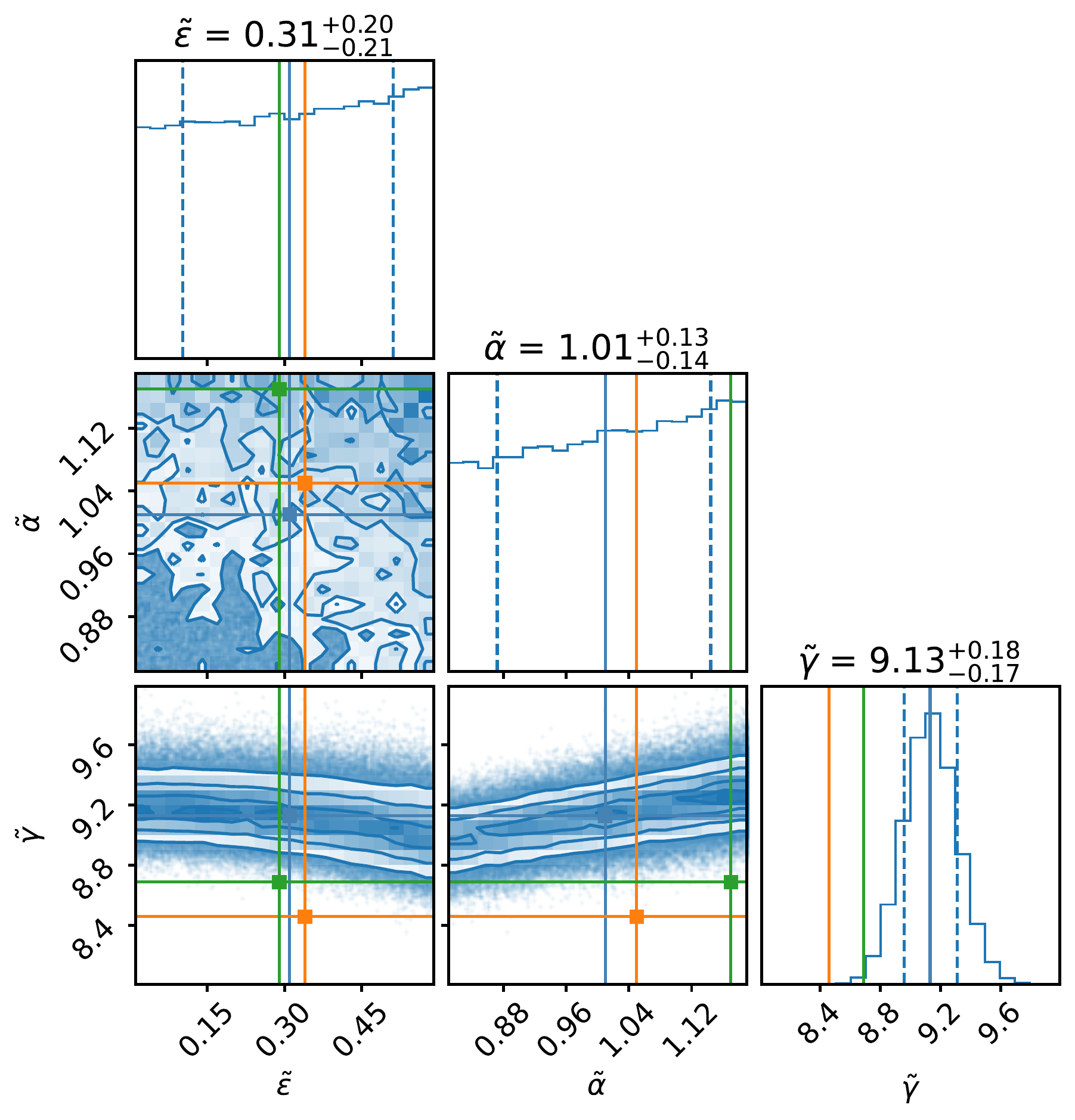}{0.45\textwidth}{(b)}}
\caption{Constraints on the parameters of the MBH--host galaxy scaling relation
(cf.~Eq.~\ref{eq:scaling}) from the GWB, with (left panel) and without (right
panel) considering the time delays between BBH coalescences and their host
galaxy mergers in the BBH population model, respectively, where  the scaling
relationship between $M\bh$ and $M\bulge$ is assumed to be redshift-independent
i.e.,  $\tiomg=0$, $\tibet=0$. The GWB is assumed to have an amplitude the same
as the common process signal discovered in the NANOGrav 12.5-year data set
\citep{Arzoumanian20cps}. The top panel in each column shows the one-dimensional
probability distribution of the parameters and the values listed above each top
panel give the median
values and $16\%$--$84\%$ quantiles of the corresponding parameter as evaluated
from the posterior distributions (also shown as the blue solid and dashed
vertical lines in the panels).  The other panels show the two-dimension
distribution of the parameters.  For comparison, the corresponding parameters of
the $M\bh$--$M\bulge$ relations taken from \citet{MM13} (orange) and
\citet{KH13} (green) are also shown in the one- and two-dimensional
distributions. See Section~\ref{sec:results:constraints} for details.}
\label{fig:gwb_limit}
\end{figure*}
\begin{figure*}[!htb]
\gridline{
\fig{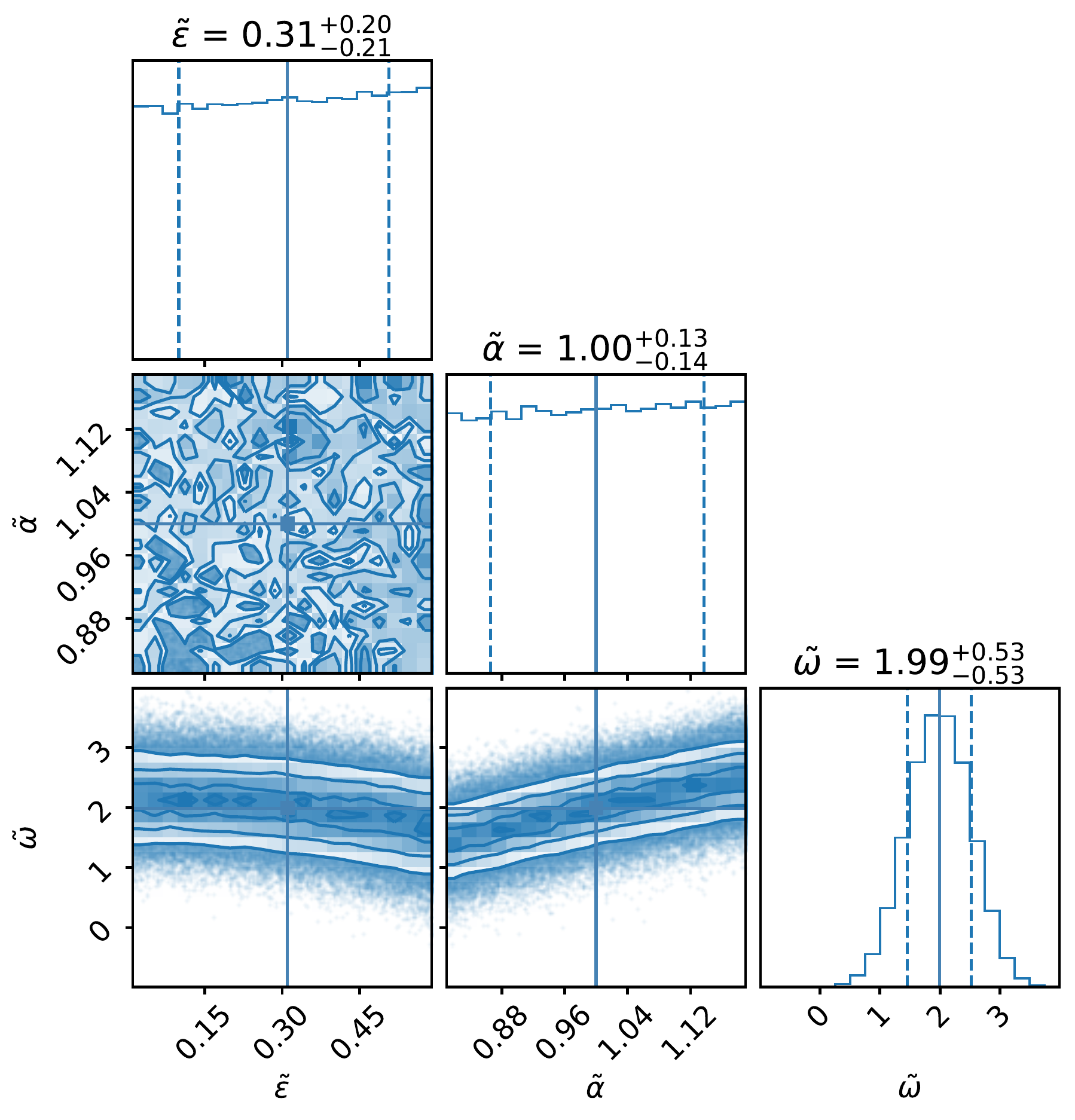}{0.45\textwidth}{(a)}
\fig{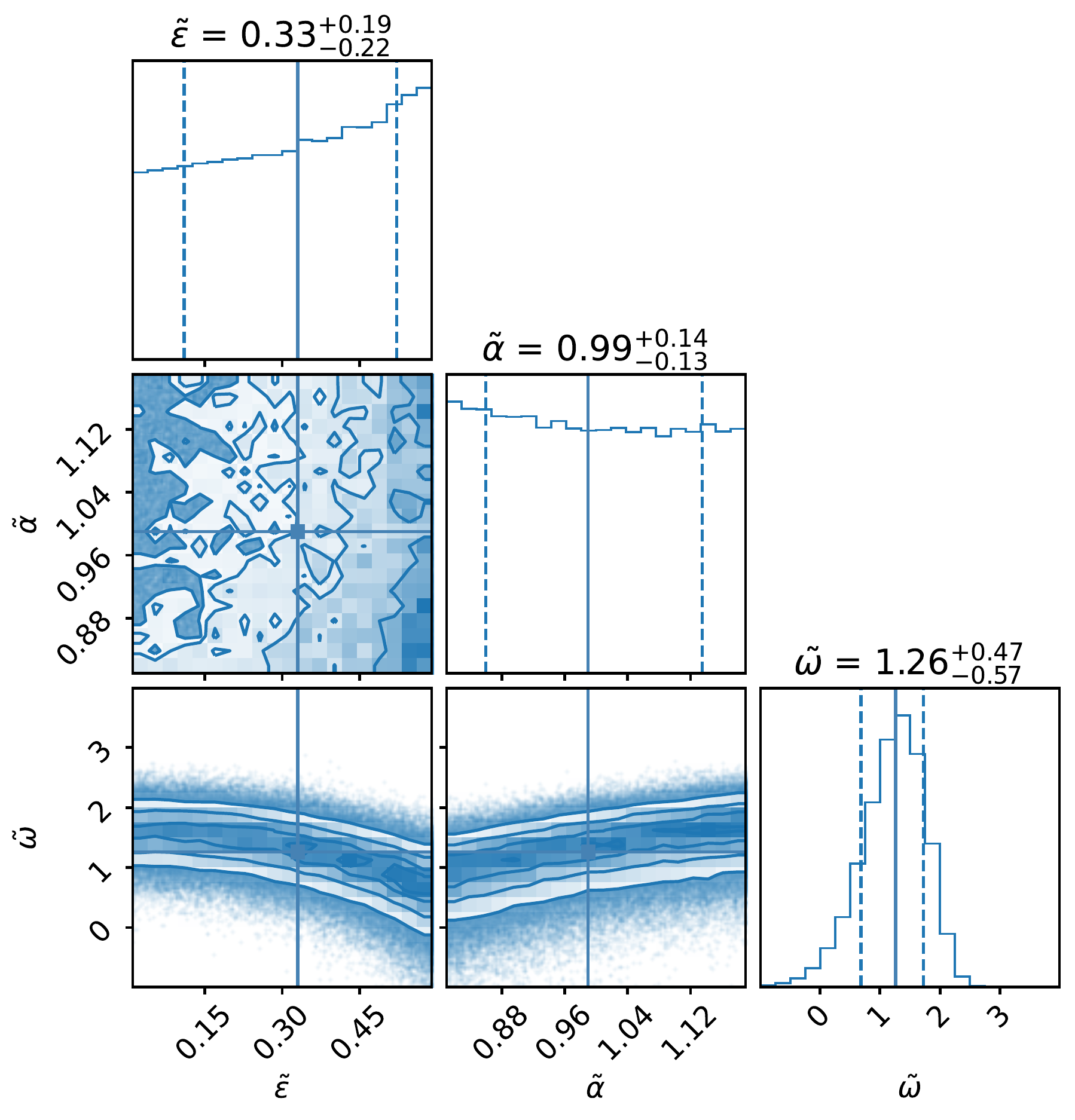}{0.45\textwidth}{(b)}}
\caption{Legends are the same as that for Figure~\ref{fig:gwb_limit}, except
that the  MBH--host galaxy scaling relation (Eq.~\ref{eq:scaling}) is set to be
redshift-dependent with $\tigam=8.69$ and $\tibet=0$,  the same as that given in
\citet{KH13}. See Section~\ref{sec:results:constraints} for details.}
\label{fig:gwb_limit_z}
\end{figure*}
\begin{deluxetable*}{lRRRCclRRRC}
\tablecaption{
Constraints on the MBH--host galaxy scaling relation from different models with
different given values of $A_{\rm yr}$.  See Section~\ref{sec:results:constraints}.  \label{tab:constraints}}
\tablehead{
\multicolumn{5}{c}{$z$-independent\tablenotemark{a}} && 
\multicolumn{5}{c}{$z$-dependent\tablenotemark{b}} \\
\cline{1-5} \cline{7-11}
\colhead{Model} & \dcolhead{\tigam} & \dcolhead{\tialp} & \dcolhead{\tieps} &
\dcolhead{\log_{10}\ayr} && 
\colhead{Model} & \dcolhead{\tiomg} & \dcolhead{\tialp} & \dcolhead{\tieps} &
\dcolhead{\log_{10}\ayr}}
\startdata
\aamaxd &  9.55^{+0.27}_{-0.26} & 1.01^{+0.13}_{-0.13} & 0.32^{+0.20}_{-0.21} & -14.72 &&  
\bbmaxd &  1.99^{+0.53}_{-0.53} & 1.00^{+0.13}_{-0.14} & 0.31^{+0.20}_{-0.21} & -14.72 \\ 
\aamidd &  8.95^{+0.20}_{-0.19} & 1.01^{+0.13}_{-0.14} & 0.31^{+0.20}_{-0.21} & -15.02 &&  
\bbmidd &  0.67^{+0.53}_{-0.60} & 1.00^{+0.14}_{-0.14} & 0.32^{+0.20}_{-0.21} & -15.02 \\
\aamind &  8.50^{+0.17}_{-0.17} & 1.00^{+0.13}_{-0.14} & 0.31^{+0.20}_{-0.21} & -15.32 && 
\bbmind & -1.00^{+0.73}_{-0.89} & 0.99^{+0.14}_{-0.13} & 0.33^{+0.19}_{-0.22} & -15.32 \\
\aamaxn &  9.13^{+0.18}_{-0.17} & 1.01^{+0.13}_{-0.14} & 0.31^{+0.20}_{-0.21} & -14.72 &&  
\bbmaxn &  1.26^{+0.47}_{-0.57} & 0.99^{+0.14}_{-0.13} & 0.33^{+0.19}_{-0.22} & -14.72 \\
\aamidn &  8.72^{+0.15}_{-0.16} & 1.01^{+0.13}_{-0.14} & 0.31^{+0.20}_{-0.21} & -15.02 && 
\bbmidn & -0.31^{+0.80}_{-1.21} & 0.99^{+0.14}_{-0.13} & 0.36^{+0.17}_{-0.24} & -15.02 \\
\aaminn &  8.35^{+0.15}_{-0.16} & 1.00^{+0.14}_{-0.14} & 0.31^{+0.20}_{-0.21} & -15.32 && 
\bbminn & -2.52^{+1.13}_{-0.99} & 1.02^{+0.12}_{-0.14} & 0.24^{+0.18}_{-0.16} & -15.32 \\
\enddata
\tablenotetext{a}{The scaling relation is assumed to be the same at different
redshifts ($z$-independent), i.e., $\tiomg=0$ and $\tibet=0$ in
Eq.~(\ref{eq:scaling}).}
\tablenotetext{b}{The scaling relation is assumed to be evolving with redshift
($z$-dependent), i.e., $\tigam=8.69$ and $\tibet=0$ in Eq.~(\ref{eq:scaling}).}
\end{deluxetable*}

\subsubsection{Redshift-independent cases}
\label{sec:zindep}

Figure~\ref{fig:gwb_limit} shows the constraints on the MBH--host galaxy scaling
relations (cf.~Eq.\,\ref{eq:scaling}) obtained by assuming that the GWB has the
same amplitude as the common process signal discovered in the NANOGrav 12.5-year
data set \citep{Arzoumanian20cps}. Here we consider the redshift-independent
scaling relation between $M\bh$ and $M\bulge$, i.e., $\tiomg=0$ and $\tibet=0$.
As seen from this figure,  $\tigam$ can be tightly constrained by the GWB
``observations''.  To produce a GWB with the same amplitude as the common process
signal, the median value of $\tigam$ needs to be $9.55$ (or $9.13$), about
$0.86-1.09{\rm\,dex}$ (or $0.44-0.67$\,dex) higher than the empirical ones
determined [e.g., $\tigam=8.46$ and $8.69$ in \citet{MM13} and \citet{KH13},
respectively]. 
The posterior distributions of
$\tigam$ show a narrow distribution with a clear peak deviating from the prior
boundaries, so that the constraint on $\tigam$  is robust.  Even though the posterior histogram of $\tigam$ in panel (a) hits the right boundary of the prior distribution,  we have checked that setting a larger boundary affects little on the constraint.
The constraints on $\tialp$ and $\tieps$ only slightly deviate from their prior
distributions,
and the constraints on them are weak.  Furthermore, there exists
some degeneracy between the constraints on the parameter $\tigam$ and the
parameter $\tialp$ or $\tieps$. 

The constraints on $\tigam$ may be understood simply by the relation between the
GWB characteristic strain amplitude $h\rmc$ and $\tigam$. At a given frequency
$f$, we divide the BBH systems into different subpopulations, with the $i$-th
subpopulation containing $N_i$ BBH systems characterized by their source-rest
chirp mass $\calM_{{\rm c},i}$ and redshift $z_i$. Then we have $h\rmc^2= \sum_i
N_i h_i^2$ with $h_i$ being the sky- and polarization-averaged strain amplitude
of the GWs emitted by a BBH system in the $i$-th subpopulation, and $h_i\propto
\calM_{{\rm c},i}^{5/3}$. On the other hand, $N_i\propto \calR_i
|\dot{f}|^{-1}\propto \calR_i \calM_{{\rm c},i}^{-5/3}$, with $\calR_i$ being
the coalescence rate of BBHs in the $i$-th subpopulation. In total, we yield
$h\rmc\propto \calR^{1/2}_i \calM_{{\rm c},i}^{5/6} \propto \calR^{1/2}_i
10^{\frac{5}{6}\tigam}$ if we keep other parameters in
Equation~(\ref{eq:scaling}) fixed,  and a larger $\tigam$ means more massive
MBHs (BBHs) in the host galaxies, and therefore leading to a larger GWB
amplitude.    For example,  according to the scaling, to make the GWB strain
amplitude larger by a factor of $2$, $\tigam$ need to be increased by
${\sim}0.36 \rmdex$ if $\calR_i$ is not significantly affected by the change of
$\tigam$ (which appears to be a reasonable approximation in the \emph{nodel}
model of Table~\ref{tab:constraints} below; see the difference of the
constrained $\tigam$ values between the {\aamaxn} model and the {\aamidn} model,
or between the {\aamidn} model and the {\aaminn} model).  

The constraints on the MBH--host galaxy scaling relations may also be affected
by the consideration of time delays between BBH coalescences,  as the time
delays affect $\calR_i$ and thus the resulting GWB strain amplitude. To make
these effects clear, we show the constraints obtained both with and without
considering the time delays in Figure~\ref{fig:gwb_limit} (see also
Tab.~\ref{tab:constraints}). By comparing the two panels of
Figure~\ref{fig:gwb_limit}, it can be seen that whether or not including the
time delays results in significantly different constraints on the scaling
relation, especially on $\tigam$. The median values of $\tigam$ is increased
by $\sim 0.15-0.42$\,dex when considering the time delays compared with those
cases without considering time delays. The reason is that the time delay effects
affect $\calR_i$ and decrease the GWB signal amplitude (see Figs.~17 and 19 of
CYL20), and thus a larger normalization for the scaling relation is required
when considering the time delays. 

The degeneracy between the constraints on the parameter $\tigam$ and the
parameter $\tialp$ or $\tieps$ can be understood in the following way.  As
mentioned before,  a larger $\tigam$ means more massive MBHs (BBHs) in the host
galaxies, and therefore leading to a larger GWB amplitude. Similarly,  a larger
$\tialp$ means a more massive MBH (BBH) in a galaxy at the high stellar-mass end
(see Eq.~\ref{eq:scaling}),  and thus also leads  to a larger GWB amplitude as
the GWB is mainly contributed by the most massive BBHs. A larger $\tieps$ also
leads to many more MBHs at the high-mass end comparing with a smaller $\tieps$
\citep[e.g.,][]{YL04qso, Lauer07bh}, and thus a larger GWB amplitude.

\subsubsection{Redshift-dependent cases}
\label{sec:zdepend}

Figure~\ref{fig:gwb_limit_z} shows the resulting posterior distributions of the
parameters $\tiomg$, $\tialp$, and $\tieps$ obtained by assuming that the GWB
amplitude is the same as the common process signal. As seen from this figure, a
strong positive dependence of the scaling relation on redshift is required
($\tiomg=1.99$ or $1.26$ in the left or right panel), suggesting that MBHs at
high redshifts are significantly larger than those inferred by the local
MBH--host galaxy scaling relation. As expected, the required redshift evolution
is more significant when considering the time delay effects,  compared with that
without considering these effects. Note that the constraints on the possible
redshift evolutions of the MBH--host galaxy scaling relation are dependent on
the model settings, especially the value of $\tigam$ in
Equation~(\ref{eq:scaling}), which may be determined by a large sample of MBHs
with dynamical mass measurements. Nevertheless, our results demonstrate that the
detection of the GWB can be used to obtain strong constraint on the potential
redshift evolutions of the MBH--host galaxy scaling relation, and hence provide
valuable information for understanding the coevolution history of MBHs and their
host galaxies.

\subsubsection{Different amplitudes of $\calA$}
\label{sec:DiffA}

We further consider some cases for which the real GWB contributed by the cosmic
BBH population is assumed to be smaller than the common process signal by a
factor of $2$ or $4$, different from the above cases in which the GWB is assumed
to be the same as the common process signal (the fiducial cases in this paper).
For these cases, we do similar analysis and obtain the ``constraints'' on the
MBH--host galaxy scaling relation (both redshift-independent and dependent) as
those for the fiducial cases described above. The resulting median values and
16\%--84\% quantiles of the parameters in Equation~(\ref{eq:scaling}) evaluated
based on the posterior distributions are shown in Table~\ref{tab:constraints}.
Below we summarize the results inferred from Table~\ref{tab:constraints}.

\begin{itemize}

\item $\tigam$ in the redshift-independent scaling relation case and $\tiomg$ in
the redshift-dependent scaling relation case are the quantities most sensitive
to both the model settings and the choices of the GWB amplitude.   
The median values of $\tialp$ and $\tieps$ change little for different
model settings and GWB amplitudes, suggesting the weak constraining power of
current ``observations'' on these two parameters.

\item If the scaling relation has no redshift evolution, the median value of
$\tigam$ decreases when the GWB amplitude is reduced. Moreover,  if the time
delays are not considered, the median values of $\tigam$ are smaller than their
counterparts in the cases where the time delays are considered.  Those are
consistent with that described in Section~\ref{sec:zindep} above.

\item If the scaling relation has redshift evolution, the median value of
$\tiomg$ decreases when the GWB amplitude is reduced. If the time delays are not
considered, the median value of $\tiomg$ are smaller than their counterparts in
the cases where the time delays are considered. 

\item The median value of $\tiomg$ decreases from a positive value to a negative
one when the GWB amplitude decreases from the fiducial value to a quarter of the
fiducial value, regardless of whether or not the time delays are considered.
This reveals the degeneracy between $\tigam$ and $\tiomg$ in the constraint. 

\end{itemize}

\subsubsection{Comparisons with previous studies}
\label{sec:compare}

Some works in the literature have also tried to obtain effective constraints on
the cosmic BBH population from the ``detection'' of or upper limits on the
stochastic GWB (e.g., \citealt{SB16, Middleton16, ChenSY19, Middleton21, IV22,
CC22}). Among these works, \citet{Middleton16, Middleton21} adopted a
parameterized function to describe the cosmic distribution of BBH mergers.
\citet{ChenSY19} explored thoroughly the constraining power of PTA detection of
the GWB or a non-detection at the $\ayr=10^{-17}$ level on a large parameter
space describing the GSMF, galaxy pair fraction and merger timescale,
$M\bh$--$M\bulge$ scaling relation as well as the eccentricities of the BBHs.
\citet{CC22} developed a quasar-based BBH model, assuming the proportionality
between the BBH population and the quasar population. We directly compare our
constraints with those obtained in \citet{SB16} and \citet{IV22} as detailed
below. However, it is not straightforward to compare our results with those by
\citet{Middleton21} and \citet{CC22} because of significantly different model
settings.

\citet{SB16} focused on constraining the MBH--host galaxy scaling relation by
using the upper limit on the GWB of $\ayr = 1\times 10^{-15}$ given by
\citet{Shannon15gwb}, which is smaller than the common process signal by a
factor of $\sim 2$. They found that the resulting constrained scaling relations
are compatible with the existing empirical ones \citep{MM13, KH13} if
considering the time delay effect, while they are not compatible with the
empirical ones if ignoring the time delay effect. Our results on the requiring
of the scaling relationship significantly different from the empirical ones are
different from theirs, which are partly due to that (1) the GWB (assuming to be
the same as the common process signal) we adopted is a factor of $\sim2$ times
larger than the upper limit they adopted, and (2) we adopt a realistic
 and comprehensive BBH population model with detailed consideration of
various underlying physics. 

\citet{IV22} estimated the GWB strain amplitude based on a semi-analytical model
allowing gas accretion during mergers to boost the MBH growth. They found that
to match the common process signal  \citep{Arzoumanian20cps}, the resulting MBH
mass function at the massive end is higher by a factor of $\sim 3$ than that
derived from observations. In our study, if assuming that the scaling relation
is redshift independent and the GWB has the same amplitude as the common process
signal, i.e., model {\aamaxd}, the scaling relation is required to have a
normalization $\tigam$ about $0.86-1.1$\,dex larger than the empirically
determined ones. By conversion, this suggests that  the MBH mass function at
$M\bh\gtrsim 10^9\msun$ in model {\aamaxd} is higher than that obtained from the
empirical relation by one order of magnitude. The constraints obtained
by \citet{IV22} are qualitatively consistent with ours, and the quantitative
difference may be due to difference in the BBH models. 

\subsection{Detections of individual BBHs}
\label{sec:results:individuals}

\begin{figure*}[!htb]
\includegraphics[width=\textwidth]{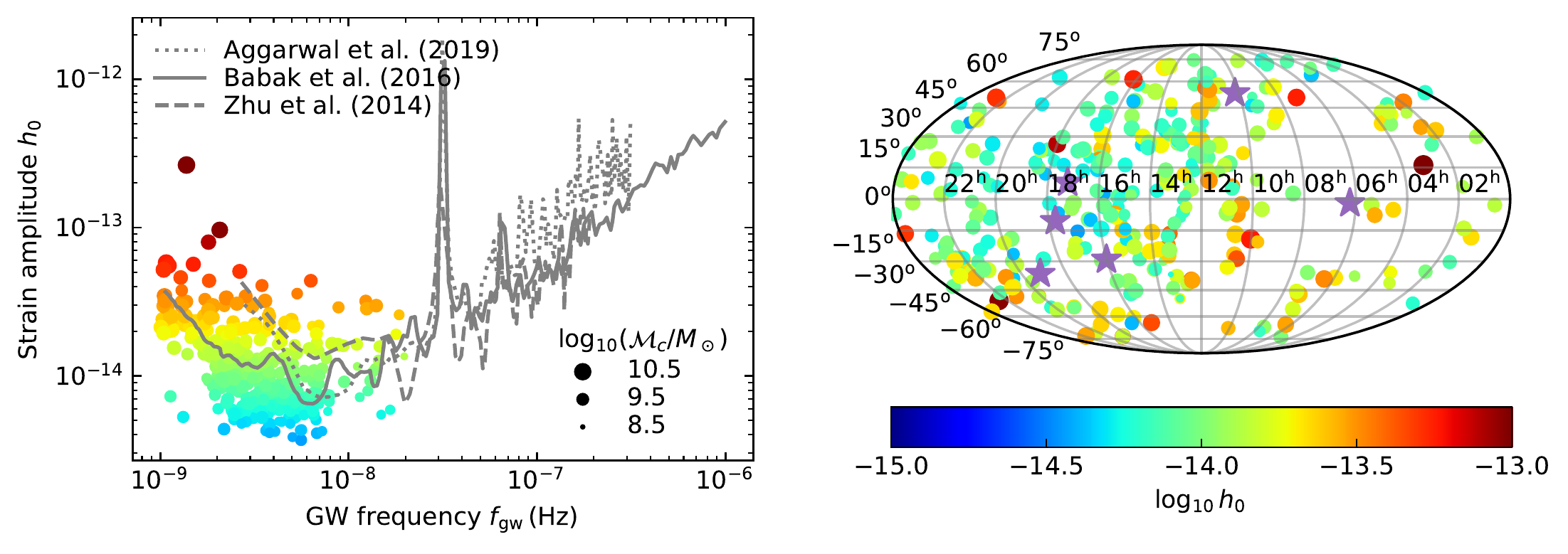}
\caption{Distributions of individual BBHs in the strain--frequency diagram (left
panel) and in the celestial sphere (right panel),  obtained from realizations of
the local BBH population based on the local sample of MBHs ($\lesssim
500$\,Mpc). To obtain those individual BBHs,  $10,000$ realizations of the local
BBH population have been conducted, based on the procedures outlined in
Section~\ref{sec:methods:individuals:local}, resulting in 347 loud sources in
344 skies being identified.  In the figure, each filled circle  located at
either its GW emission frequency (left panel) or its celestial coordinates
(right panel) corresponds to a BBH system with GW strain amplitude
(cf.~Eq.~\ref{eq:h0}) being above the 95\% upper limit skymap taken from
\citet{Mingarelli17}, where the color and size of the circle indicates the GW
strain amplitude and the BBH chirp mass, respectively. The BBH population model
adopted here produces a stochastic GWB with amplitude the same as the fiducial
value of the common process signal \citep{Arzoumanian20cps}, i.e.,  the
{\aamaxd} model in Tab.~\ref{tab:constraints}.  In the left panel, the
sky-averaged 95\% upper limit on the strain amplitude of individual BBHs based
on the current PTAs, i.e., NANOGrav 11-year timing data \citep{Aggarwal19cgw},
EPTA DR1 timing data \citep{Babak16cgw}, and PPTA DR1 timing data
\citep{ZhuXingjiang14cgw}, are also shown as references by the dotted, solid and
dashed curves, respectively. In the right panel, the six best MSPs among the
EPTA pulsars are shown by the purple star symbols \citep{Babak16cgw}. See
Section~\ref{sec:results:individuals:tng} for details.}
\label{fig:loud_occu}
\end{figure*}
\begin{figure*}[!htb]
\includegraphics[width=\textwidth]{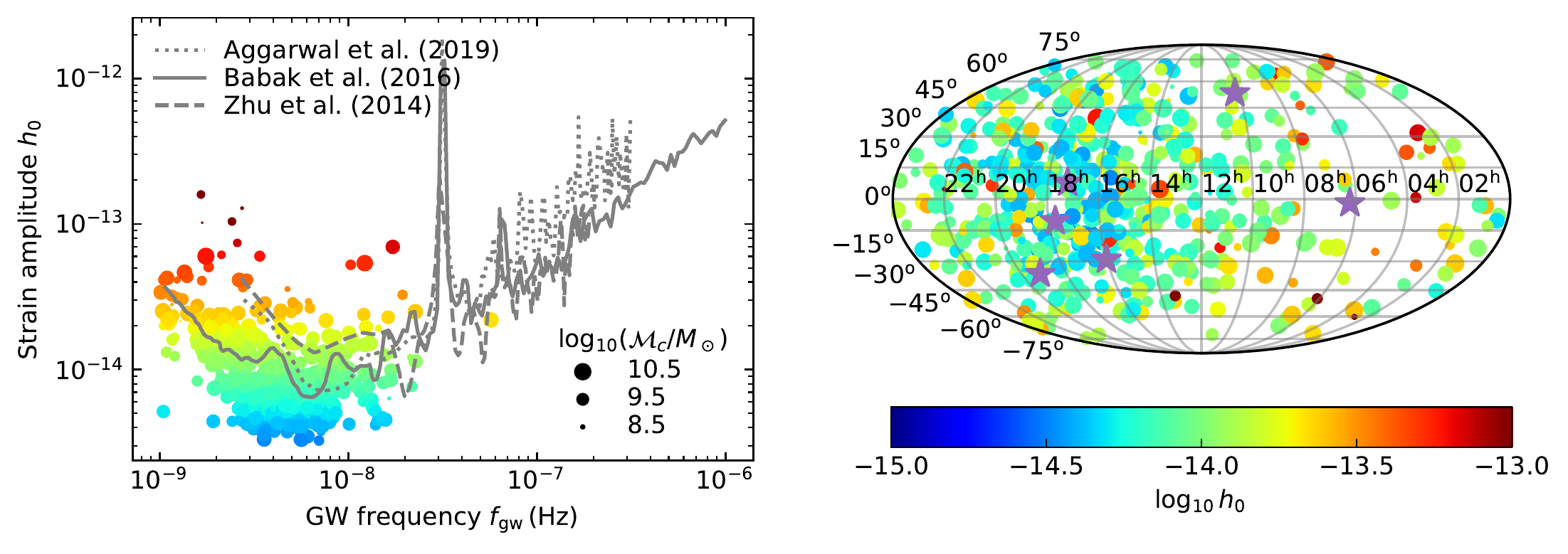}
\caption{Legends are the same as that for Figure~\ref{fig:loud_occu},  except
that the global BBH population are plotted here.  Among the $10,000$
realizations of the BBH global populations, $587$ loud individual BBHs in $563$
skies are identified with strain amplitudes above the EPTA sensitivity skymap
\citep{Mingarelli17}, as represented by each filled circle in both panels. See
Section~\ref{sec:results:individuals:tng} for more details.}
\label{fig:loud_indv}
\end{figure*}

With the BBH population model constrained by the GWB ``observations'', we now
proceed to estimate the occurrence rate of individual BBHs and confront them
with the sensitivity curves of some ongoing and planned PTAs to investigate
their detectability.   

\subsubsection{Detection prospects for the Current PTAs}
\label{sec:results:individuals:tcg}

Figure~\ref{fig:loud_occu} shows the distributions of individual ``detectable''
BBHs in the strain-frequency diagram and in the celestial sphere, obtained from
realizations of the local population of BBHs by the procedures described  in
Section~\ref{sec:methods:individuals:local}.  The BBH population model adopted
here is the \aamaxd model (see Tab.~\ref{tab:constraints}), which leads to a GWB
with strain amplitude the same as the common process signal
\citep{Arzoumanian20cps}.  Among the $10,000$ realizations conducted, $347$ loud
individual BBHs inside $344$ skies are identified with the strain amplitude
$h_0$ at the corresponding celestial coordinates (see the right panel of
Fig.~\ref{fig:loud_occu}) and GW frequencies above the 95\% upper limit skymap
of individual BBHs for EPTA taken from \citet{Mingarelli17}.  We define these
sources as ``detectable'' ones by the EPTA (as a representative of the current
PTAs), which are plotted as color filled circles in the figure.  For the $347$
detectable sources, the maximum and median luminosity distances are $\sim
422\mpc$ and $\sim 99\mpc$, respectively, and the majority of them have
luminosity distances being within $\sim 200\mpc$. As  also seen from
Figure~\ref{fig:loud_occu},  for individual CGW sources, those most massive BBH
systems (e.g., with chirp mass $>10^9\msun$) tend to be detected first. 

Our estimate on the detection probability is larger than that obtained in
\citet{Mingarelli17} (i.e., $131$ skies containing loud individual BBHs among
$75,000$ realizations) by a factor of $\sim$20. Both the larger sample size of
the local galaxies and the larger normalization ($\tigam$) of the adopted
MBH--host galaxy scaling relation are responsible for the larger detection
probability, especially the latter one. However, the predicted detection
probability is still too small for the currently operative PTAs. 

We also plot these $347$ ``detectable'' sources and the sky-averaged 95\% upper
limits on the strain amplitude of individual BBHs set by existing PTAs in the
left panel of Figure~\ref{fig:loud_occu} \citep{Aggarwal19cgw, Babak16cgw,
ZhuXingjiang14cgw}.  A larger fraction of these sources are below the
sky-averaged upper limits of the PTAs,  which indicates that the detection
probabilities are boosted when the angular sensitivities of the PTAs are
accounted for especially  in the regime of small detection probabilities (see
also Fig.~1 of \citealt{Mingarelli17}).  

Figure~\ref{fig:loud_indv} shows the detection prospects of the global
population of BBHs as individual CGW sources targeted by those existing PTAs.
The adopted BBH population model is the same as that adopted in
Figure~\ref{fig:loud_occu} (i.e., {\aamaxd}, see Tab.~\ref{tab:constraints}). To
get the detection statistics, we also conduct $10,000$ realizations of the
global BBH population, among which $587$ loud individual BBHs belonging to $563$
skies are identified with $h_0$ at the corresponding sky locations and GW
frequencies above the $95\%$ upper limit skymap taken from \citet{Mingarelli17}.
The detection probability for the global BBH population is only mildly larger
than that resulting from the local BBH population  (by a factor $\la 2$),
suggesting that the detection capabilities of the current PTAs on individual
BBHs are still limited to relatively local volumes.

In the right panels of Figures~\ref{fig:loud_occu} and \ref{fig:loud_indv},  the
six best millisecond pulsars among the EPTA pulsar set are shown by purple stars
\citep{Babak16cgw}. The ``detectable'' loud BBHs tend to accumulate around those
best millisecond pulsars, though their sky locations have been isotropically
generated in the realizations (see also \citealt{Mingarelli17}).

\begin{figure*}[!htb]
\centering
\includegraphics[width=1.0\textwidth]{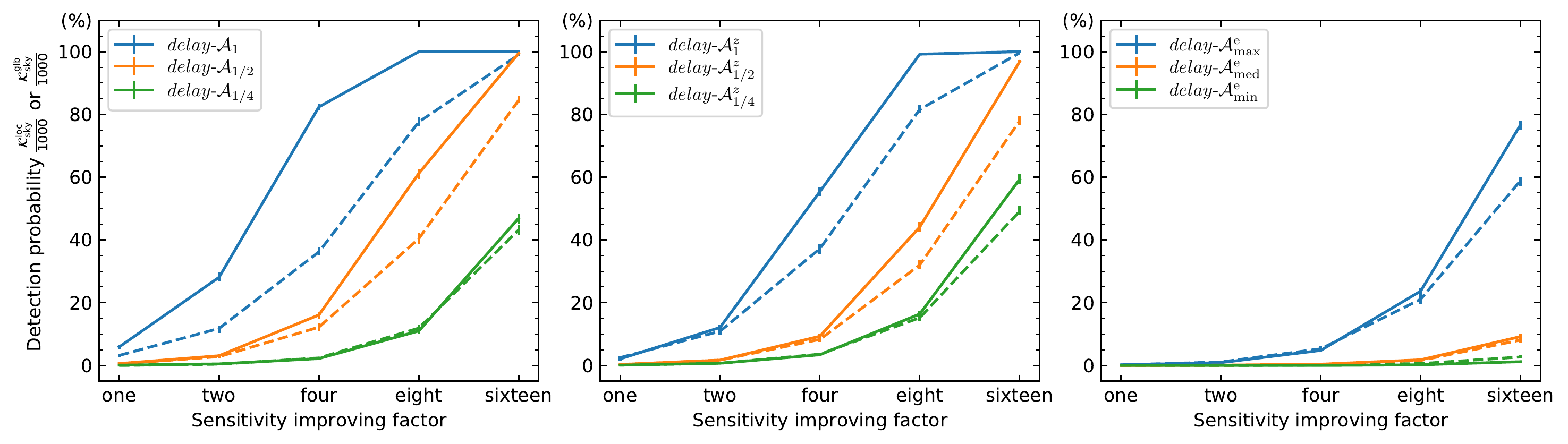}
\caption{
Detection probabilities of individual BBHs for different sets of models and for
different PTAs capabilities. We consider the current PTA detection capability by
adopting the 95\% upper limit skymap taken from \citet{Mingarelli17}, and future
PTA capabilities with sensitivity a factor of $1/2$, $1/4$, $1/8$ or $1/16$ of
the current one. The results for models adopting the constrained scaling
relations with and without redshift evolution are shown in the left and middle
panels, respectively; while the right panel shows the results obtained from
those models adopting the empirical scaling relations. The dashed and solid
curves correspond to detections of the local and global BBH populations,
respectively. The data points and their error bars are adapted from
Table~\ref{tab:tcg}.
}
\label{fig:plot_Table4}
\end{figure*}

Tables~\ref{tab:tcg} lists the expected numbers that characterize the detection
probabilities of individual BBHs for different sets of models and for different
PTAs. As for the models, we consider not only those newly constrained scaling
relations in Section~\ref{sec:results:constraints} (see
Tab.~\ref{tab:constraints}), but also those admitting the empirically determined
ones in the literature (see Tab.~\ref{tab:empirical}) for comparisons.  As for
the PTA capabilities, we adopt the 95\% upper limit skymap taken from
\citet{Mingarelli17} as a representative of the current PTA detection
capability, and those by improving the \citet{Mingarelli17} sensitivity skymap
by a factor of $2$, $4$, $8$, or $16$, as the representatives for the evolution
of the detection capability. 
Here the sensitivity being improved by a factor of $X$ means the skymap
being scaled downward by a factor of $X$.
For each model, $\klocsky$ and $\klocsrc$ represent the number of skies
containing at least one loud individual source and the total number of loud
individual sources in $1,000$ realizations of the local BBH population,
respectively; so are $\kglbsky$ and $\kglbsrc$,  except that they are for the
global BBH population. For each of the quantities, the median value and
$16\%$--$84\%$ quantiles are listed in the Table.
For clarity,  the detection probabilities for different sets of models and for different
PTA detection capabilities are also summarized in Figure~\ref{fig:plot_Table4}.

As seen from Table~\ref{tab:tcg} and Figure~\ref{fig:plot_Table4},  the
sensitivities of the current PTAs may be not sufficient for the detection of
individual BBHs, unless they are improved by a factor of close to $10$ or more.
For example, adopting the model {\aamaxd}, the current sensitivity skymap only
has a detection probability of 3.2\% if limited to the local BBH population, and
it increases to 5.9\% if adopting the global population. To get a detection
probability greater than $95\%$, the current sensitivity skymap needs to be
improved by a factor of $\sim 8$, in which case the detection probability
reaches $100\%$ and the number of individual BBHs that  can be detected is about
$9$. If limited to the local BBH population, it needs to be improved by a factor
of $16$, under which circumstance the detection probability can reach $99\%$ and
the expected detection number is reduced about $4$--$5$. In this case,  the
maximum luminosity distance of the detectable BBHs is ${\sim} 480\mpc$, and the
median luminosity distance is  ${\sim} 90\mpc$. We emphasize that the model
adopted here is the quite optimistic one since the scaling relation in the model
has a normalization being greater than that of those empirical ones by nearly
one order of magnitude (see Tab.~\ref{tab:constraints}). 

As seen from Table~\ref{tab:tcg} and Figure~\ref{fig:plot_Table4},   the results
obtained from the model {\bbmaxd} are slightly different from those described
above by adopting the model  {\aamaxd}. Although the models {\aamaxd} and
{\bbmaxd} produce the same GWB strain amplitude,  the former requires more
massive BBH systems in the local volume, while the latter requires a larger
contribution from relatively higher redshifts. When the PTA detection capability
is limited to be relatively local,  the model {\aamaxd} yields a larger
detection probability of individual BBHs. Note that in the {\bbmaxd} model,
adopting the current sensitivity skymap results in a slightly higher detection
probability of the local BBH population compared to that of the global
population (the left points of the blue curves in the middle panels of Fig.~\ref{fig:plot_Table4} ),  which is due to that the two populations are obtained in different
ways as described in Sections~\ref{sec:misc:model} and \ref{sec:misc:sample}. 
For the similar reason,  some other models (e.g.,  {\ccmind}; the green curves in the
right panel of Fig.~\ref{fig:plot_Table4})  also result in higher detection
probability of the local BBH population compared with that of the global ones. 

As seen from Table~\ref{tab:tcg} and Figure~\ref{fig:plot_Table4}, for detection
of the BBH population,  in the redshift-dependent model the current sensitivity
skymap needs to be improved by a factor of $8$ so that the detection probability
for the global BBH population can reach $99.2\%$ and the expected detection
number is  $\sim 5$. If the sensitivity skymap is improved  by a factor of $16$,
the detection probability for the local BBH population reaches $99.6\%$ and the
expected detection number is  $\sim 6$. Note that the sensitivity can be
improved in different ways,  as Section~\ref{sec:misc:sns} shows that at a given
frequency the threshold sensitivity on the strain amplitude of individual
sources $h_{0,{\rm th}}$ is proportional to different variable: $\sigma_a$,
$T^{-1/2}$, $\Delta t^{1/2}$, and $\calN\psr^{-1/2}$ (see Eq.~\ref{eq:sns4cgw}).
For example, if we improve the timing precision by a factor of $16$ (i.e.,
reducing $\sigma_a$ by a factor of $16$), the sensitivity can be improved by a
factor of $16$; or if we double both the observational period $T$ and the
cadence $1/\Delta t$, then the same improved sensitivity can be achieved by
reducing $\sigma_a$ by a factor of $8$.

As seen from Table~\ref{tab:tcg} and Figure~\ref{fig:plot_Table4}, in the
{\ccmaxd} model, the resulting detection probabilities are significantly lower
mainly due to the significant smaller $\tigam$ in this model than that in the
{\aamaxd} model. Quantitatively, the expected detection probability is merely
$0.2\%$, for both the local and global BBH populations, according to the current
sensitivity skymap for EPTA. Even the sensitivity is improved by a factor of
$16$, the detection probability can only reach $58.9\%$ for the local BBH
population and $76.7\%$ for the global BBH population, both of which are smaller
than $95\%$. 

\subsubsection{Detection prospects for CPTA and SKAPTA}
\label{sec:results:individuals:tng}

\begin{figure*}[!htb]
\centering
\includegraphics[width=1.0\textwidth]{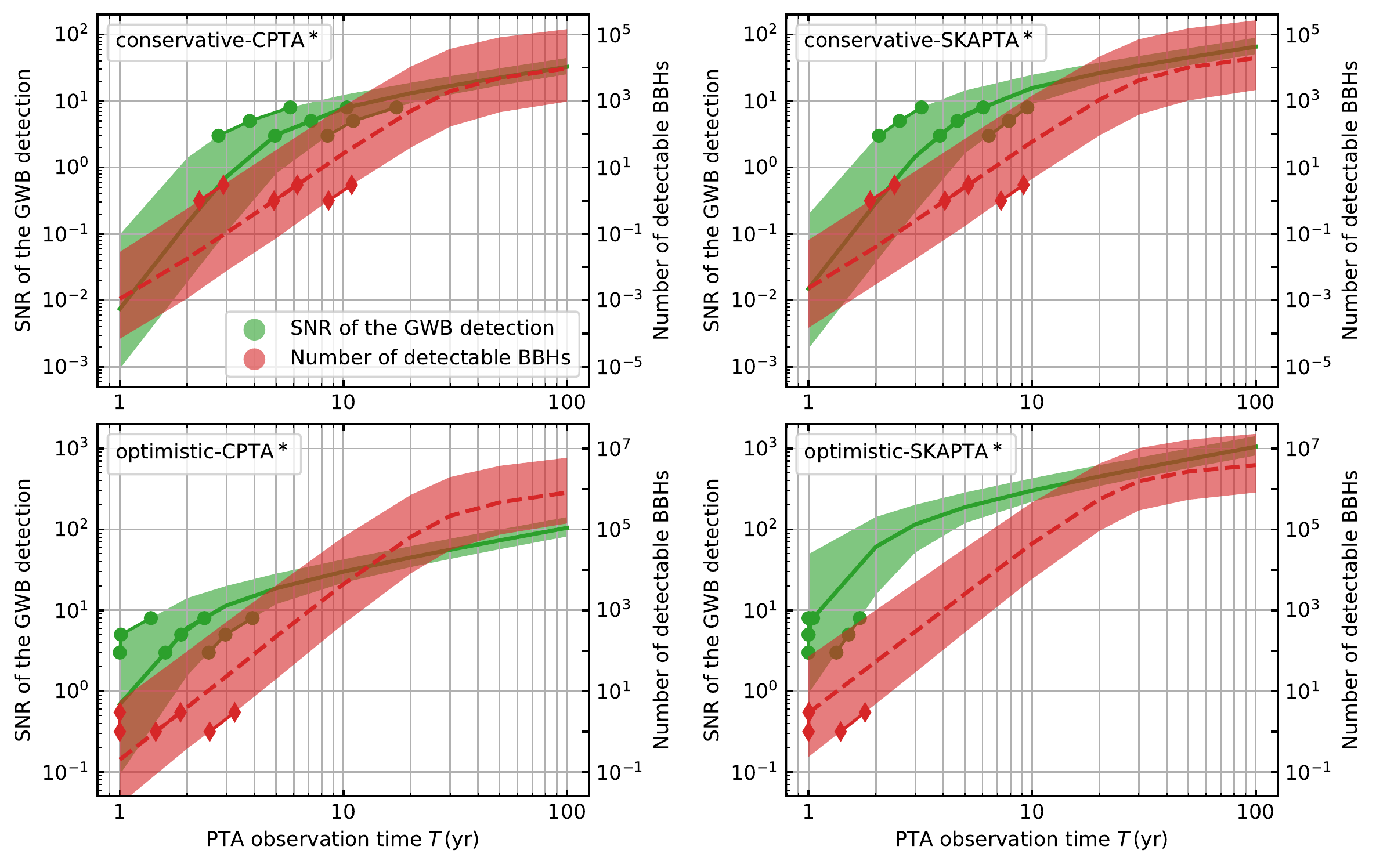}
\caption{Expected SNRs of the stochastic GWB detection (green; left $y$-axis)
and expected numbers of detectable BBHs (red; right $y$-axis) by different PTAs
as a function of PTA observation time $T$.  In each panel,  the `$\ast$' symbol
appended to each PTA name indicates that  the assumed parameters are the same as
those listed in Tab.~\ref{tab:PTAs},  except that the PTA observation time $T$
is set as a free parameter.   To quantify the uncertainties of the expected
quantities, we have considered three sets of models, i.e., {\aamaxd}, {\ccmaxd}
and {\ccmidd} (see Tab.~\ref{tab:tng}), respectively, corresponding to the upper
boundary, thick curve in the middle,  and the lower boundary of each shaded
region. Top panels show the results for the two conservative PTA configurations,
and the lower panels for the two optimistic PTA configurations. In each panel,
the green filled circles mark the observation times when the SNR of the
stochastic GWB detection equals the values of $3$, $5$ and $8$,
and the red filled diamonds mark the observation time when the expected number
of detectable BBHs equals the values of $1$ and $3$
(with a threshold SNR $\rho_{\rm th}=3$).  See
Section~\ref{sec:results:individuals:tng} for details.}
\label{fig:joint_trends}
\end{figure*}

\begin{figure*}[!htb]
\gridline{
\fig{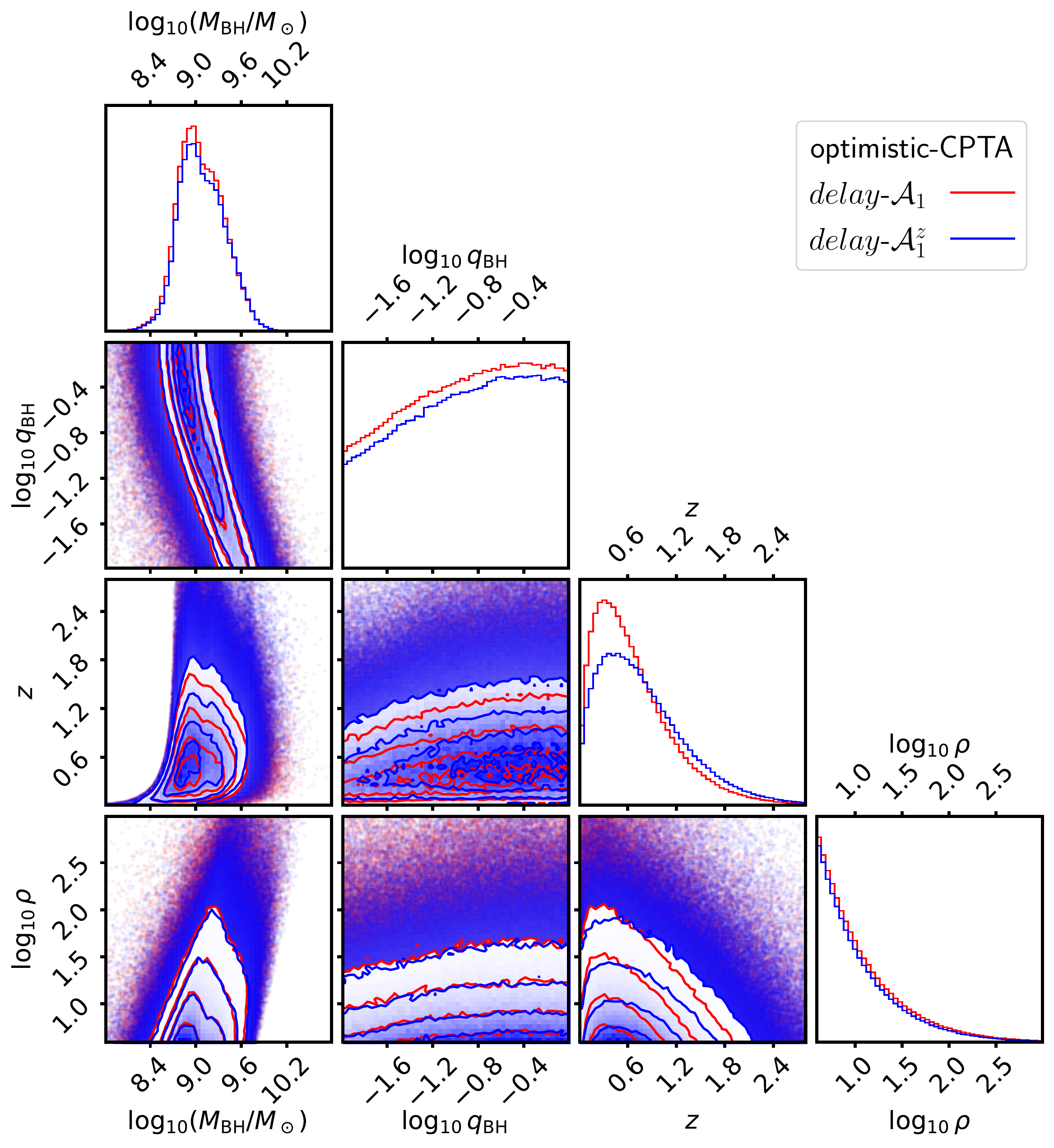}{0.45\textwidth}{(a)}
\fig{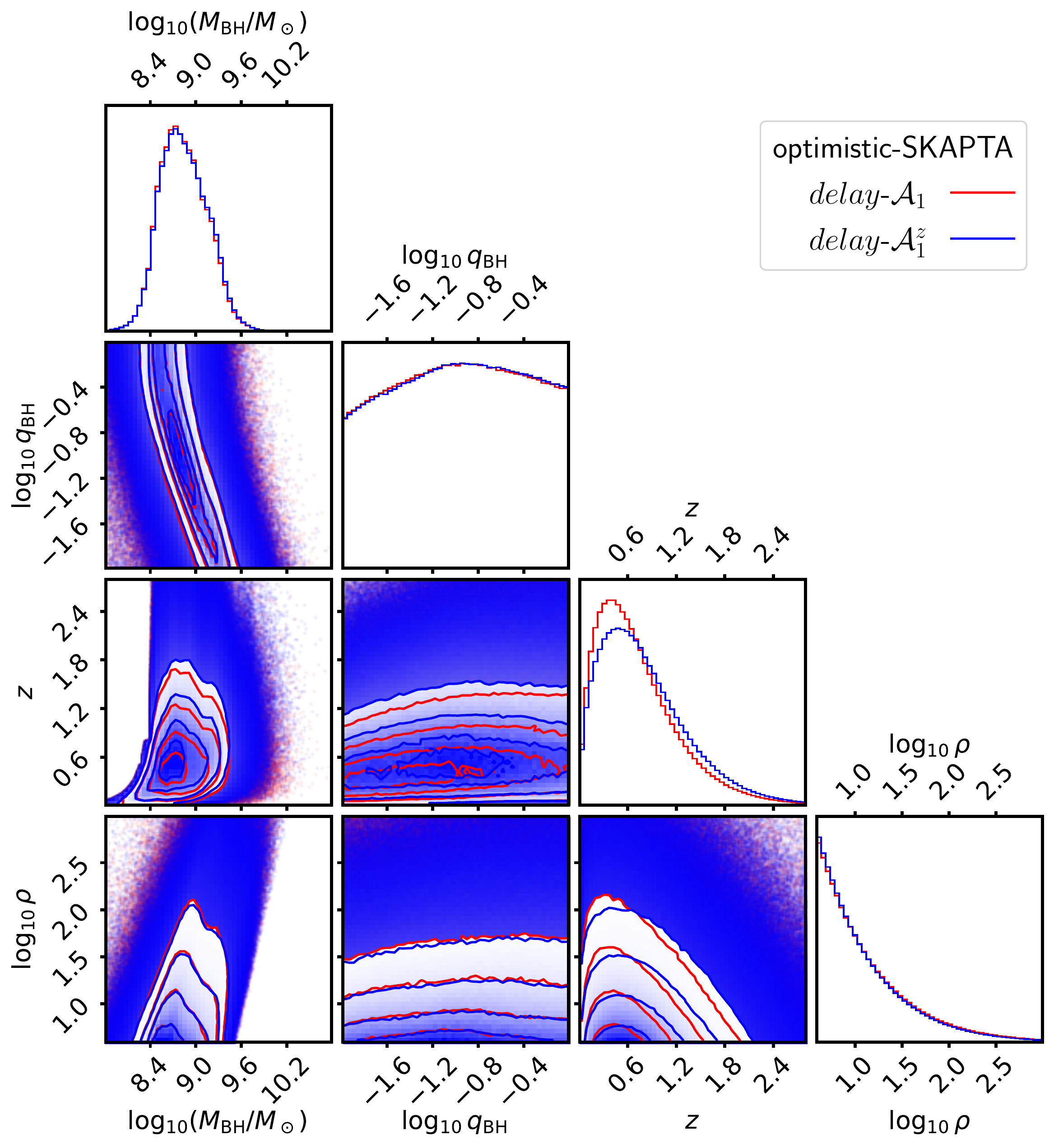}{0.45\textwidth}{(b)}}
\caption{Distributions of the individual BBHs that are detectable by the
optimistic-CPTA (left) and the optimistic-SKAPTA (right),  in the source
parameter space of the total mass $M\bh$, mass ratio $q\bh$, redshift $z$, and
detection SNR $\rho$.  The red and blue colors in the figure correspond to the
{\aamaxd} and {\bbmaxd} models, respectively.  For both models,  the sources
from one single realization of the global BBH population are shown in the
figure. The one-dimensional histograms in the top panels show the number
distributions of the detectable sources. This figure shows that the different
models, i.e., with and without redshift evolution, result in different redshift distributions of
the detectable BBHs (e.g., different peak heights and relative number ratios of low-z to high-z sources).  
 See Section~\ref{sec:results:individuals:tng} for details.}
\label{fig:hist_loud_scaling}
\end{figure*}

\begin{figure*}[!htb]
\gridline{
\fig{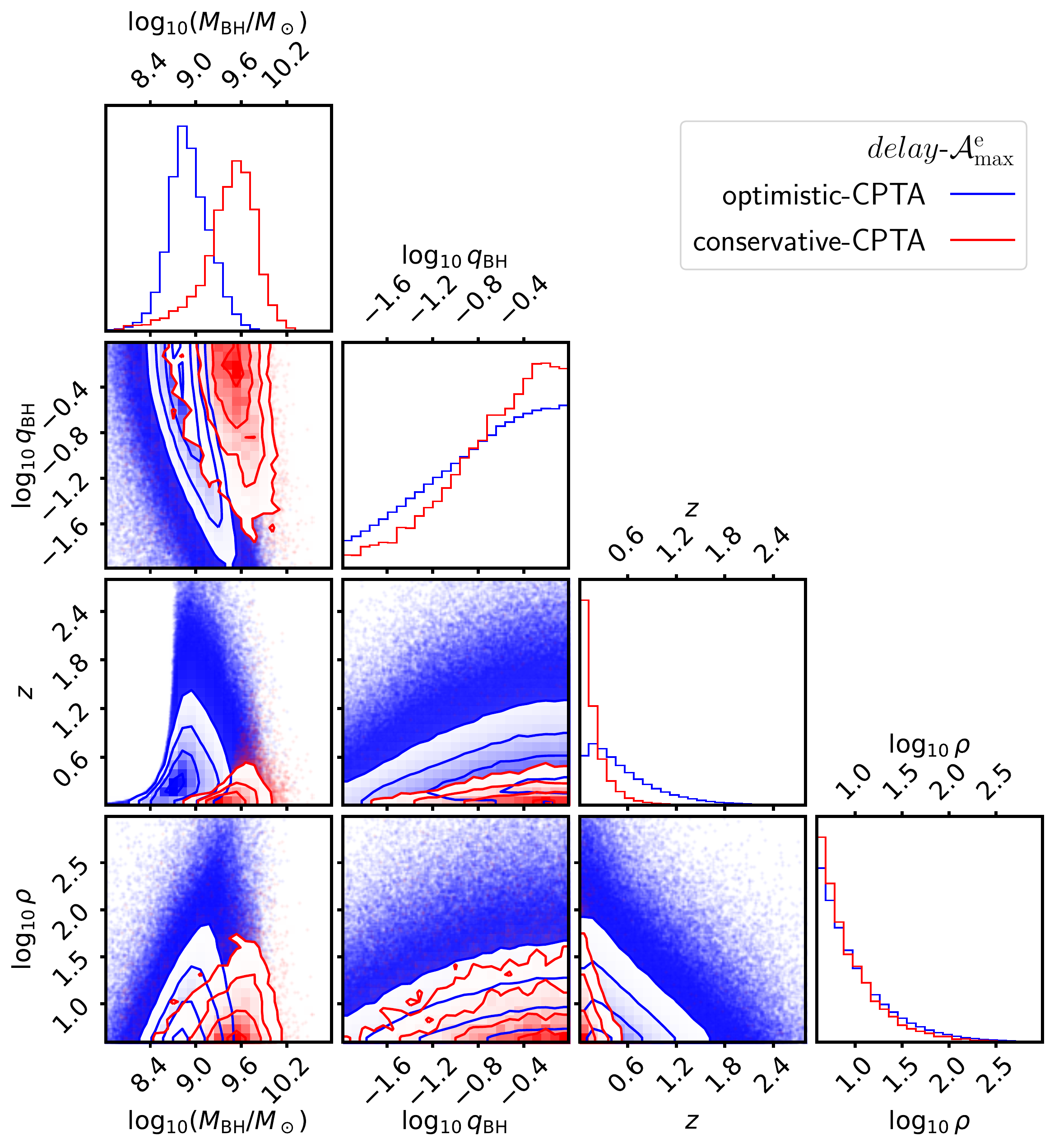}{0.45\textwidth}{(a)} 
\fig{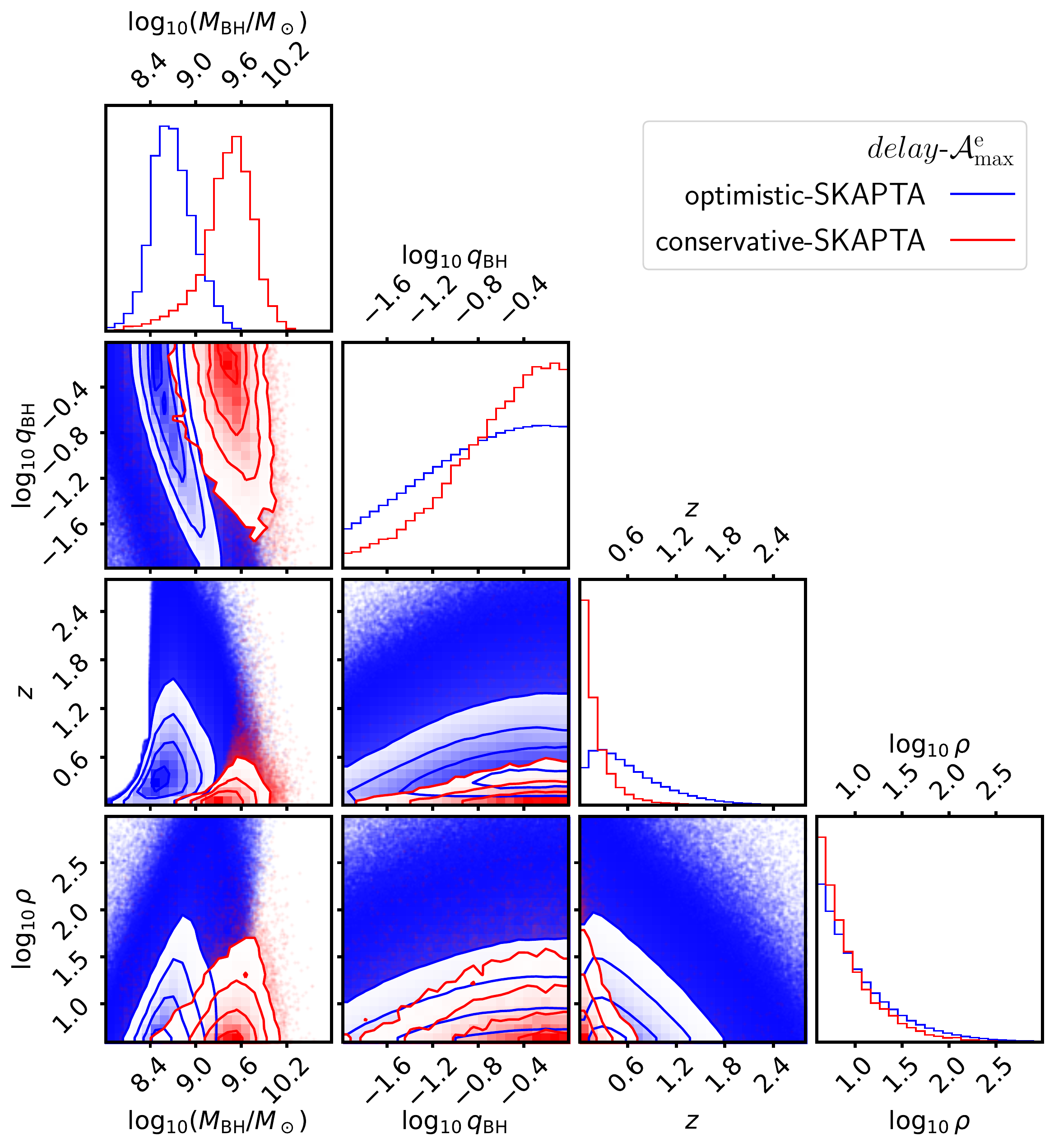}{0.45\textwidth}{(b)}}
\caption{Distributions of the individual BBHs that are detectable by the two
CPTA configurations (left) and by the two SKAPTA configurations (right),  in the
source parameter space of the total mass $M\bh$, mass ratio $q\bh$, redshift
$z$,  and detection SNR $\rho$. In both panels,  the  ${\ccmaxd}$ model is
adopted.  The blue and red colors correspond to the optimistic PTA
configurations and the conservative PTA configurations, respectively. As for the
optimistic PTA configurations, the detectable sources from $10$ realizations of
the global BBH population are shown,  and for the conservative PTA
configurations, the detectable sources from $10,000$ realizations of the global
BBH population are shown. The one-dimensional histograms in the top panels show
the number distributions of the detectable sources. This figure shows that the
small sample of the BBHs detected at the beginning tend to have larger masses
(peaking at $\sim 4\times10^9M_\odot$), larger mass ratios (close to $1$), and
lower redshifts ($\sim 0.1$), compared with the large BBH populations to be
detected by the optimistic PTAs. See Section~\ref{sec:results:individuals:tng}
for details.}
\label{fig:hist_loud_config}
\end{figure*}

We proceed to investigate the detection prospects of individual BBHs by CPTA and
SKAPTA \citep{NanRendong11, LeeKJ16,  Smits09}  (see Tab.~\ref{tab:PTAs}),  and
the results are shown in Table~\ref{tab:tng},  similarly as done in
Table~\ref{tab:tcg}, and also in
Figures~\ref{fig:joint_trends}-\ref{fig:hist_loud_config}. For the conservative
configurations of CPTA and SKAPTA,  we use the same variables ($\klocsky$,
$\klocsrc$,  $\kglbsky$, and $\kglbsrc$) as those listed in Table~\ref{tab:tcg}.
However, for the optimistic configurations of both PTAs, the detection
probabilities are high enough that the results from a single realization can
give robust statistics. Therefore, we use $\nlocsky$ and $\nlocsrc$ to represent
the number of skies containing at least one loud individual source and the total
number of loud individual sources in a single realization of the local BBH
population. Similarly, we use $\nglbsky$ and $\nglbsrc$ to represent the similar
quantities, but for the global population. The mean values of the above
quantities averaged over $100$ realizations are listed in Table~\ref{tab:tng}.

As seen from Table~\ref{tab:tng},  if adopting the conservative CPTA/SKAPTA
configurations with $T=5\yr$, a positive detection of individual BBHs can be
achieved if the stochastic GWB has an amplitude close to the reported common
process signal, and it can also be achieved for all other models listed in
Table~\ref{tab:tng} if adopting the optimistic CPTA/SKAPTA configurations with
$T=20\yr$ (Tab.~\ref{tab:PTAs}). For example,  given the sensitivities of the
conservative-CPTA and the conservative-SKAPTA,  the detection probabilities of
the global population of BBHs are expected to be both $100\%$ in the  {\aamaxd}
model,    and $98.2\%$ and $100\%$, respectively, in the model {\aamidd}. If the
model {\bbmaxd} is adopted, the detection probabilities are expected to be both
$100\%$ for the two conservative PTA configurations, while if the model
{\bbmidd} is adopted, the two detection probabilities are $93.0\%$ and $99.9\%$,
respectively. The detection probabilities reduce to $66.8\%$ and $92.1\%$,
respectively, if adopting the model {\ccmaxd}. If adopting the optimistic CPTA
configuration and the model {\ccmind}, which produces a GWB with the smallest
amplitude among those models considered in this paper, one may still expect to
detect $182$ and $798$ BBHs for the local and global BBH populations,
respectively. 


Figure~\ref{fig:joint_trends} shows the expected number of detectable BBHs and
the expected SNR of the GWB by different PTAs as a function of the observation
time $T$,  where the PTA configurations are the same as those listed in
Table~\ref{tab:PTAs} except that $T$ is taken as a free parameter.  This figure
illustrates how the detection probability/number of individual BBHs and the SNR
for the GWB detection depend on the BBH population model and the PTA detection
capabilities. The SNRs for the GWB detection is estimated by using
Equation~(\ref{eq:snr4gwb}),  the SNRs of individual BBHs are estimated by
Equation~\eqref{eq:snr4cgw},  and the threshold sensitivities on the strain
amplitude of individual sources are obtained by using
Equation~\eqref{eq:sns4cgw}.  Because both the detection probability/number of
individual BBHs and the SNR for the GWB detection depend on the BBH population
model, we choose the following three different models to quantify their
uncertainties: {\aamaxd},  {\ccmaxd}, and {\ccmidd}. The {\aamaxd} model and the
{\ccmidd} model set the upper  and lower boundaries of both quantities as shown
by the shaded regions in Figure~\ref{fig:joint_trends}, respectively. The
results obtained for the two conservative PTA configurations and two optimistic
PTA configurations are shown in the top and bottom panels, respectively. In each
panel of the figure, we use the green filled circles to mark the observation
times when the SNR of the GWB detection equals $3$, $5$, and $8$, respectively.
While we use the red filled diamonds to mark the times when the expected number
of detectable BBHs reaches $1$ and $3$ (with a threshold SNR $\rho_{\rm th}=3$),
respectively. 

As seen from Figure~\ref{fig:joint_trends},  the detections of the stochastic
GWB and individual BBHs are expected to be realized at the time close to each
other for those PTAs shown in the figure (see green circles and red diamonds).
Taking the conservative-CPTA$^\ast$ as an example, the GWB is expected to be
detected with an SNR of $3$ after an observation time of $2.8/4.9/8.5\yr$ when
the model {\aamaxd}/{\ccmaxd}/{\ccmidd} is adopted. The corresponding
observation time needed for a detection number of $3$ for individual BBHs
(corresponding to a detection probability of $\sim 95\%$, see
Sec.~\ref{sec:misc:sns}) is $2.9/6.2/10.9\yr$. If increasing the detection SNR
to $5$ (or $8$), the stochastic GWB is expected to be detected in
$3.8/7.2/11.1\yr$ (or $5.8/10.4/17.3\yr$),  and the time needed for a detection
number of $3$ for individual BBHs is $3.3/7.1/12.9\yr$ (or $3.7/8.0/14.8\yr$).
When the conservative-SKAPTA$^\ast$ is adopted, the detection of the GWB with an
SNR of $3$ (or $5$ and $8$) is expected in an observation time of
$2.1/3.9/6.4\yr$ (or $2.6/4.6/7.9\yr$ and $3.2/6.0/9.5\yr$); and the detection
of individual BBHs with $95\%$ detection probability is expected  in
$2.4/5.2/9.2\yr$ (or $2.8/5.9/10.7\yr$ and $3.1/6.7/12.1\yr$). If the optimistic
PTA configurations are considered, positive detections are expected in a shorter
time.  For the optimistic-CPTA$^\ast$, both detections can be achieved in $\sim
1-3$ years,  and for the optimistic-SKAPTA$^\ast$, both detections can be
achieved in $\sim 1-2$ years.

Furthermore, the breakthrough of the individual BBH detection is expected to be
followed by a quick cumulation of a sizable sample of BBHs as shown by the red
shaded region in Figure~\ref{fig:joint_trends}.  For example,  the number of
individual BBHs detectable by CPTA with the conservative configuration  can
amount to $\sim 100$ in $\sim 5-18$ years since the first detection of
individual BBHs.  For the optimistic configuration of SKAPTA,  one can have
$\sim 100$ individual BBH detections in $\sim 1-3$ years since the first
detection of individual BBHs.
We note again here that when the PTA sensitivity is sufficiently high, there
may be many individual BBHs in a single frequency bin that have characteristic
strain higher than the sensitivity curve, and there might be a limit on the
total number of these individual BBHs that can be extracted from a single
frequency bin. It is quite important to develop efficient data analysis
methods  for this as those done for a similar problem, i.e.,  extracting
individual double white dwarfs in the low-frequency band for Laser
Interferometer Space Antenna (LISA) \citep{Cornish03, Mohanty06, Littenberg11,
ZhangXH21}.

The distributions of the properties of the detected BBHs contain critical
information of the underlying population, and thus may be used to distinguish
different BBH population models. Figure~\ref{fig:hist_loud_scaling} shows the
total mass ($M\bh$), mass ratio ($q\bh$), redshift ($z$), and detection SNR
($\rho$) distributions of those individual BBHs detectable by the
optimistic-CPTA and the optimistic-SKAPTA in the left and right panels,
respectively. In this figure, the red and blue colors represent the models
{\aamaxd} and {\bbmaxd}, respectively. As seen from the left and right panels,
the detectable individual sources resulting from the two models have similar
distributions in the total mass $M\bh$, mass ratio $q\bh$, and detection SNR
$\rho$, although the absolute number of detectable sources in the {\aamaxd}
model is a little larger than that in the  {\bbmaxd} model. The main difference
occurs in the redshift distribution.  If the scaling relation has redshift
evolution, there are less detectable sources at the low-redshifts (i.e.,
$z\lesssim 1$) while more detectable sources at the high-redshifts (i.e.,
$1\lesssim z\lesssim 2.4$). By comparing the left and the right panels, one
observes that the peak distributions of $M\bh$ and $q\bh$ move towards smaller
values, because the optimistic-SKAPTA has a substantially higher sensitivity
than the optimistic-CPTA does.

Figure~\ref{fig:hist_loud_config} shows the comparison of the individual sources
detectable by the conservative PTAs and the optimistic PTAs. As seen in the
figure,  the small sample of the BBHs detected by PTAs at the beginning do not
follow the bulk distributions of the large samples to be accumulated later:
they tend to have larger masses (peaking at $\sim 4\times10^9M_\odot$), larger
mass ratios (close to $1$), and lower redshifts ($\sim 0.1$), compared with the
large BBH populations to be detected by the optimistic PTAs.

\section{Conclusions}
\label{sec:conclusions}

In this paper, we conduct a joint study of the two types of GW signals in the
PTA band, including both the stochastic GWB and individual CGW sources,
originating from the cosmic population of BBHs. We adopt the formation and
evolution model of the cosmic BBHs from CYL20, which is composed of several
astrophysical ingredients, including the GSMF, the MRPG, and the MBH--host
galaxy scaling relation. First, we set constraints on the BBH population model
with the recently discovered common process signal by several different PTAs
\citep{Arzoumanian20cps, Goncharov21cps, ChenSiyuan21cps, Antoniadis22cps},
assuming that the stochastic GWB has a strain amplitude the same as or a
fraction of the common process signal. Among the model ingredients, we focus on
the constraints on the MBH--host galaxy scaling relation, leaving the other
ingredients fixed to their fiducial choices set by current available
observations. Both the redshift independent and redshift-dependent scaling
relations are considered, the latter may contain important information about the
coevolution of MBHs with their host galaxies. Second, with both the constrained
scaling relations and those empirical ones from the literature, we explore the
detection prospects of individual BBHs by both the PTAs that have been operated
for many years and those new and/or future ones using more powerful radio
telescopes.  Third,  we expect that the BBH model can be further constrained
when the expected detection probabilities/numbers of the individual sources and
the parameter distributions of these detectable sources are confronted with real
detections in the future.  Below we summarize the main conclusions of this
study.

\begin{itemize}

\item If the MBH-host galaxy scaling relation has no redshift evolution, in
order to produce a stochastic GWB with the same amplitude as the common process
signal reported by \citet{Arzoumanian20cps}, i.e., $\ayr= 1.92\times 10^{-15}$,
the median value of $\tigam$ (i.e., the normalization factor) in the scaling
relation (Eq.~\ref{eq:scaling}) needs to be $9.55$. The constrained value is
about $0.86$--$1.1\rmdex$ higher than the ones empirically determined in the
$M\bh$--$M\bulge$ relations taken from \citet{KH13} and \citet{MM13}. If the
scaling relation is redshift-dependent, then the stochastic GWB can be produced
with the same amplitude as the common process signal by introducing significant
redshift evolution to the scaling relation. We demonstrate that the GWB
observation can be used to set strong constraints on the potential redshift
evolution of the MBH--host galaxy scaling relation, therefore provide valuable
information about the coevolution of MBHs and their host galaxies.

\item For PTAs with the current EPTA sensitivity,  the detection probability of
individual BBHs is low for all the models considered in this paper.  For the
model with those empirically determined MBH--host galaxy scaling relations, the
detection probabilities of individual BBHs by the currently operative PTAs
(e.g., \citealt{Aggarwal19cgw, Babak16cgw, ZhuXingjiang14cgw}) are negligible.
For example, with the $95\%$ upper limit skymap of EPTA taken from
\citet{Mingarelli17}, the detection probability of individual BBHs is merely
$0.2\%$, if we adopt the $M\bh$--$M\bulge$ relation from \citet{KH13} in the
model. However, with the constrained MBH--host galaxy scaling relations by the
reported common process signal, the detection probability increases
considerably, e.g., being boosted to $3.2\%$-$5.9\%$. Despite this significant
increase, the detection probabilities are still too small to guarantee a
positive detection of individual BBHs by current PTAs.

\item The detection prospect of individual BBHs is promising,  if the
sensitivity of PTAs can be improved by an order of magnitude compared with the
current EPTA sensitivity.  With the redshift-independent scaling relation
constrained by the reported common process signal, the detection probability of
individual sources reaches $100\%$ if the current EPTA sensitivity skymap is
improved by a factor of $8$, with about $9$ detections being expected. If the
detectable BBHs are limited to the local volume ($\lesssim 500$\,Mpc), the
detection probability reaches $99.0\%$ among our realizations if the current
EPTA sensitivity skymap is improved by a factor of $16$, with $4$--$5$
detections being expected. If adopting the constrained scaling relation with
redshift evolution, the resulting detection probabilities and detection
statistics of individual BBHs are somewhat smaller than those from the above
case without redshift evolution. In addition, adopting the empirically
determined scaling relations such as the $M\bh$--$M\bulge$ relation from
\citet{KH13}, the detection probability can only reach $58.9\%$ for the local
BBH population and $76.7\%$ for the global BBH population even if the current
EPTA sensitivity skymap could be improved by a factor of $16$.

\item For CPTA and SKAPTA with the assumed configurations (Tab.~\ref{tab:PTAs}),
we estimate both the expected SNR of the GWB detection and the expected
detection probability/number of individual BBHs, as a function of the PTA
observation time, by considering a wide range of models. For these PTA
configurations, it is expected that the detections of the GWB and individual
BBHs come at the time close to each other. For the conservative configuration of
CPTA,  the GWB is expected to be detected  with an SNR of $3$ within an
observation period of ${\sim}3$--$9\yr$. The corresponding time needed for the
detection of individual BBHs with a detection probability of $95\%$ is
${\sim}3$--$11\yr$. If increasing the detection SNR to 5 (or 8), the stochastic
GWB is expected to be detected in ${\sim}4$--$12\yr$ (or ${\sim}6$--$18\yr$),
while individual BBHs are expected to be detected with a probability of $95\%$
in ${\sim}4$--$13\yr$ (or ${\sim}4$--$15\yr$). We emphasize here that both the
GWB and individual BBHs can be detected by CPTA within an observation period of
$3-4$ years, if the contribution from cosmic population of BBHs to the GWB is
the same as the common process signal. For the conservative configuration of
SKAPTA,  the required time for the GWB detection with an SNR of $3$ ($5$,  or
$8$) is ${\sim}3$--$7\yr$ (${\sim}3$--$8\yr$ or ${\sim}4$--$10\yr$); and the
required times for the detection of individual BBHs with an SNR threshold of $3$
($5$, or $8$) with $95\%$ detection probability is ${\sim}3$--$10\yr$
(${\sim}3$--$11\yr$,  or ${\sim}4$--$13\yr$). For the optimistic configurations
of CPTA and SKAPTA, detections of both signals are expected in $\sim 1-3$ and
$1-2$ years, respectively. 

\item The breakthrough of individual BBH detection is expected to be followed by
a quick accumulation of a sizable sample of individual sources (e.g.,  the
number of detectable BBHs increases by two orders of magnitude in $\sim 1-12$
years since the first detection depends on the PTA configurations). We
demonstrate that different BBH population models are expected to be effectively
distinguished as they lead to different parameter distributions of detectable
sources.  Our study shows that the possible redshift evolution of the MBH--host
galaxy scaling relation can be constrained by the redshift distribution of the
detectable individual BBHs. A positive redshift evolution of the scaling
relation means less low-redshift sources and more high-redshift ones than the
model in which the scaling relation has no redshift evolution but producing the
same GWB strain amplitude. 

\end{itemize}

\section*{acknowledgements}
This work is partly supported by the National SKA Program of China and the
National Key Program for Science and Technology Research and Development (grant
No. 2020SKA0120101,  2022YFC2205201,  2020YFC2201400),  the National Natural
Science Foundation of China (grant nos. 11721303,  12173001,  11673001,
11991052), and the Strategic Priority Program of the Chinese Academy of
Sciences (grant no. XDB 23040100).
\centerwidetable
\begin{longrotatetable}
\begin{deluxetable*}{lRRRRcclRRRRcclRRRR}
\tabletypesize{\scriptsize}
%
\tablecaption{
Detection prospects of individual BBHs by EPTA and its scaled sensitivities.
\label{tab:tcg}}
%
%
\tablehead{ \multicolumn{5}{c}{$z$-independent} &&&
\multicolumn{5}{c}{$z$-dependent} &&& \multicolumn{5}{c}{Empirical relations} \\
\cline{1-5} \cline{8-12} \cline{15-19}
\colhead{Model} & \dcolhead{\klocsky} & \dcolhead{\klocsrc} &
\dcolhead{\kglbsky} & \dcolhead{\kglbsrc} &&& \colhead{Model} &
\dcolhead{\klocsky} & \dcolhead{\klocsrc} & \dcolhead{\kglbsky} &
\dcolhead{\kglbsrc} &&& \colhead{Model} & \dcolhead{\klocsky} &
\dcolhead{\klocsrc} & \dcolhead{\kglbsky} & \dcolhead{\kglbsrc}}
\startdata
\multicolumn{19}{c}{ {Sensitivity skymap of EPTA}}\\ \hline
\aamaxd & 32^{+4}_{-6} & 32^{+5}_{-6} & 59^{+5}_{-7} & 62^{+4}_{-8} &&& \bbmaxd
& 25^{+6}_{-7} & 25^{+6}_{-6} & 21^{+6}_{-6} & 21^{+6}_{-5} &&& \ccmaxd &
2^{+1}_{-1} & 2^{+1}_{-1} & 2^{+1}_{-1} & 2^{+1}_{-1} \\ 
\aamidd & 5^{+3}_{-2} & 5^{+3}_{-2} & 6^{+3}_{-2} & 6^{+3}_{-2} &&& \bbmidd &
3^{+2}_{-2} & 3^{+2}_{-2} & 3^{+3}_{-1} & 3^{+3}_{-1} &&& \ccmidd & 0^{+0}_{-0}
& 0^{+0}_{-0} & 0^{+1}_{-0} & 0^{+1}_{-0} \\
\aamind & 0^{+1}_{-0} & 0^{+1}_{-0} & 1^{+1}_{-1} & 1^{+1}_{-1} &&& \bbmind &
1^{+1}_{-1} & 1^{+1}_{-1} & 1^{+2}_{-1} & 1^{+2}_{-1} &&& \ccmind & 0^{+0}_{-0}
& 0^{+0}_{-0} & 0^{+0}_{-0} & 0^{+0}_{-0} \\
\aamaxn & 41^{+6}_{-7} & 43^{+6}_{-8} & 47^{+7}_{-8} & 48^{+7}_{-8} &&& \bbmaxn
& 5^{+3}_{-2} & 5^{+4}_{-2} & 8^{+2}_{-3} & 8^{+3}_{-3} &&& \ccmaxn &
6^{+2}_{-2} & 6^{+2}_{-2} & 4^{+2}_{-2} & 4^{+2}_{-2} \\
\aamidn & 6^{+3}_{-2} & 6^{+3}_{-2} & 4^{+2}_{-2} & 4^{+2}_{-2} &&& \bbmidn &
5^{+2}_{-2} & 5^{+2}_{-2} & 5^{+2}_{-2} & 5^{+3}_{-2} &&& \ccmidn & 0^{+0}_{-0}
& 0^{+0}_{-0} & 0^{+1}_{-0} & 0^{+1}_{-0} \\
\aaminn & 0^{+1}_{-0} & 0^{+1}_{-0} & 1^{+1}_{-1} & 1^{+1}_{-1} &&& \bbminn &
4^{+2}_{-2} & 4^{+3}_{-2} & 3^{+2}_{-2} & 3^{+2}_{-2} &&& \ccminn & 0^{+0}_{-0}
& 0^{+0}_{-0} & 0^{+0}_{-0} & 0^{+0}_{-0} \\ \hline
\multicolumn{19}{c}{ {Sensitivity skymap of EPTA scaled downward by a factor of
2}}\\ \hline
\aamaxd & 118^{+9}_{-14} & 125^{+11}_{-15} & 281^{+15}_{-13} & 326^{+25}_{-14}
&&& \bbmaxd & 109^{+9}_{-10} & 115^{+10}_{-10} & 121^{+10}_{-9} &
129^{+13}_{-10} &&& \ccmaxd & 11^{+4}_{-3} & 11^{+4}_{-3} & 9^{+4}_{-2} &
9^{+4}_{-2} \\ 
\aamidd & 28^{+5}_{-6} & 28^{+5}_{-6} & 31^{+4}_{-5} & 32^{+4}_{-5} &&& \bbmidd
& 17^{+4}_{-4} & 17^{+5}_{-4} & 17^{+4}_{-3} & 18^{+4}_{-4} &&& \ccmidd &
0^{+1}_{-0} & 0^{+1}_{-0} & 1^{+1}_{-1} & 1^{+1}_{-1} \\
\aamind & 4^{+2}_{-2} & 4^{+2}_{-2} & 5^{+2}_{-2} & 5^{+2}_{-2} &&& \bbmind &
7^{+2}_{-2} & 7^{+2}_{-2} & 7^{+2}_{-2} & 7^{+2}_{-2} &&& \ccmind & 0^{+0}_{-0}
& 0^{+0}_{-0} & 0^{+0}_{-0} & 0^{+0}_{-0} \\
\aamaxn & 148^{+10}_{-9} & 161^{+11}_{-12} & 236^{+13}_{-12} & 269^{+17}_{-15}
&&& \bbmaxn & 28^{+4}_{-6} & 28^{+4}_{-5} & 58^{+7}_{-7} & 59^{+9}_{-8} &&&
\ccmaxn & 30^{+8}_{-5} & 31^{+7}_{-6} & 21^{+5}_{-3} & 22^{+5}_{-4} \\
\aamidn & 31^{+6}_{-5} & 32^{+6}_{-6} & 24^{+5}_{-3} & 24^{+5}_{-3} &&& \bbmidn
& 28^{+5}_{-6} & 28^{+5}_{-5} & 26^{+5}_{-5} & 27^{+4}_{-6} &&& \ccmidn &
1^{+1}_{-1} & 1^{+1}_{-1} & 1^{+1}_{-1} & 1^{+1}_{-1} \\
\aaminn & 4^{+2}_{-2} & 4^{+2}_{-2} & 4^{+2}_{-2} & 4^{+2}_{-2} &&& \bbminn &
25^{+4}_{-6} & 25^{+5}_{-6} & 12^{+5}_{-3} & 12^{+5}_{-3} &&& \ccminn &
0^{+1}_{-0} & 0^{+1}_{-0} & 0^{+0}_{-0} & 0^{+0}_{-0}  \\ \hline
\multicolumn{19}{c}{ {Sensitivity skymap of EPTA scaled downward by a factor of
4}}\\ \hline
\aamaxd & 362^{+14}_{-11} & 452^{+16}_{-23} & 824^{+11}_{-10} & 1752^{+29}_{-51}
&&& \bbmaxd & 372^{+15}_{-16} & 465^{+20}_{-23} & 554^{+14}_{-14} &
804^{+29}_{-26} &&& \ccmaxd & 53^{+9}_{-7} & 54^{+11}_{-6} & 48^{+6}_{-6} &
49^{+7}_{-7} \\ 
\aamidd & 122^{+14}_{-12} & 131^{+13}_{-15} & 161^{+11}_{-11} & 175^{+13}_{-13}
&&& \bbmidd & 83^{+8}_{-8} & 87^{+7}_{-9} & 93^{+9}_{-10} & 97^{+10}_{-10} &&&
\ccmidd & 3^{+2}_{-2} & 3^{+2}_{-2} & 4^{+1}_{-2} & 4^{+1}_{-2} \\
\aamind & 24^{+4}_{-6} & 24^{+4}_{-5} & 22^{+4}_{-4} & 22^{+4}_{-4} &&& \bbmind
& 36^{+5}_{-6} & 36^{+7}_{-6} & 34^{+7}_{-5} & 34^{+7}_{-5} &&& \ccmind &
1^{+1}_{-1} & 1^{+1}_{-1} & 0^{+1}_{-0} & 0^{+1}_{-0} \\
\aamaxn & 424^{+15}_{-19} & 548^{+24}_{-20} & 770^{+13}_{-15} & 1465^{+42}_{-35}
&&& \bbmaxn & 121^{+8}_{-13} & 128^{+10}_{-13} & 360^{+16}_{-20} &
445^{+26}_{-23} &&& \ccmaxn & 128^{+9}_{-12} & 136^{+13}_{-13} & 118^{+15}_{-10}
& 126^{+14}_{-12} \\
\aamidn & 132^{+12}_{-12} & 143^{+13}_{-15} & 131^{+10}_{-11} & 141^{+11}_{-14}
&&& \bbmidn & 120^{+9}_{-9} & 127^{+10}_{-9} & 129^{+11}_{-9} & 140^{+12}_{-12}
&&& \ccmidn & 6^{+2}_{-3} & 6^{+2}_{-3} & 6^{+2}_{-2} & 6^{+2}_{-2} \\
\aaminn & 26^{+5}_{-7} & 26^{+5}_{-6} & 17^{+3}_{-4} & 17^{+3}_{-4} &&& \bbminn
& 115^{+12}_{-10} & 122^{+12}_{-11} & 60^{+8}_{-8} & 62^{+9}_{-8} &&& \ccminn &
2^{+2}_{-2} & 2^{+2}_{-2} & 1^{+1}_{-1} & 1^{+1}_{-1} \\ \hline
\multicolumn{19}{c}{ {Sensitivity skymap of EPTA scaled downward by a factor of
8}}\\ \hline
\aamaxd & 776^{+14}_{-13} & 1492^{+35}_{-30} & 1000^{+0}_{-0} &
9014^{+93}_{-111} &&& \bbmaxd & 816^{+14}_{-11} & 1701^{+44}_{-40} &
992^{+2}_{-4} & 4781^{+70}_{-72} &&& \ccmaxd & 211^{+11}_{-16} & 235^{+13}_{-17}
& 237^{+9}_{-17} & 269^{+14}_{-19} \\ 
\aamidd & 404^{+18}_{-16} & 518^{+25}_{-27} & 612^{+15}_{-17} & 946^{+32}_{-33}
&&& \bbmidd & 320^{+16}_{-12} & 385^{+22}_{-16} & 441^{+16}_{-15} &
582^{+26}_{-25} &&& \ccmidd & 16^{+4}_{-4} & 16^{+4}_{-4} & 18^{+4}_{-3} &
18^{+4}_{-3} \\
\aamind & 119^{+10}_{-11} & 127^{+10}_{-13} & 110^{+9}_{-10} & 116^{+11}_{-10}
&&& \bbmind & 152^{+11}_{-9} & 166^{+12}_{-12} & 164^{+10}_{-12} &
179^{+10}_{-14} &&& \ccmind & 6^{+2}_{-2} & 6^{+2}_{-2} & 2^{+2}_{-1} &
2^{+2}_{-1} \\
\aamaxn & 821^{+10}_{-12} & 1712^{+26}_{-41} & 1000^{+0}_{-1} &
7553^{+79}_{-101} &&& \bbmaxn & 394^{+11}_{-15} & 502^{+18}_{-19} &
954^{+7}_{-7} & 3103^{+40}_{-61} &&& \ccmaxn & 400^{+18}_{-16} & 515^{+17}_{-24}
& 511^{+13}_{-17} & 716^{+24}_{-36} \\
\aamidn & 429^{+18}_{-15} & 561^{+22}_{-25} & 552^{+14}_{-17} & 801^{+27}_{-34}
&&& \bbmidn & 393^{+14}_{-15} & 498^{+20}_{-21} & 528^{+16}_{-13} &
753^{+26}_{-28} &&& \ccmidn & 29^{+6}_{-4} & 30^{+6}_{-5} & 31^{+5}_{-4} &
32^{+5}_{-5} \\
\aaminn & 120^{+10}_{-8} & 129^{+9}_{-13} & 88^{+7}_{-10} & 92^{+9}_{-11} &&&
\bbminn & 405^{+12}_{-16} & 518^{+22}_{-25} & 257^{+12}_{-13} & 296^{+16}_{-13}
&&& \ccminn & 10^{+2}_{-3} & 10^{+2}_{-3} & 4^{+2}_{-1} & 4^{+2}_{-1} \\ \hline
\multicolumn{19}{c}{ {Sensitivity skymap of EPTA scaled downward by a factor of
16}}\\ \hline
\aamaxd & 990^{+3}_{-4} & 4577^{+60}_{-60} & 1000^{+0}_{-0} &
44206^{+179}_{-243} &&& \bbmaxd & 996^{+2}_{-1} & 5698^{+69}_{-72} &
1000^{+0}_{-0} & 26368^{+182}_{-135} &&& \ccmaxd & 589^{+12}_{-17} &
885^{+32}_{-29} & 767^{+13}_{-16} & 1440^{+41}_{-33} \\ 
\aamidd & 845^{+14}_{-9} & 1884^{+35}_{-51} & 995^{+1}_{-3} & 5146^{+67}_{-71}
&&& \bbmidd & 779^{+16}_{-12} & 1515^{+43}_{-45} & 968^{+6}_{-5} &
3451^{+59}_{-62} &&& \ccmidd & 81^{+8}_{-9} & 84^{+10}_{-10} & 92^{+8}_{-9} &
96^{+10}_{-9} \\
\aamind & 432^{+17}_{-16} & 565^{+23}_{-24} & 469^{+15}_{-18} & 633^{+24}_{-26}
&&& \bbmind & 491^{+17}_{-11} & 680^{+17}_{-24} & 593^{+17}_{-16} &
900^{+30}_{-31} &&& \ccmind & 27^{+5}_{-5} & 28^{+4}_{-6} & 12^{+4}_{-4} &
12^{+4}_{-4} \\
\aamaxn & 993^{+3}_{-2} & 4880^{+50}_{-60} & 1000^{+0}_{-0} &
36718^{+161}_{-156} &&& \bbmaxn & 832^{+12}_{-9} & 1781^{+41}_{-35} &
1000^{+0}_{-0} & 19333^{+131}_{-165} &&& \ccmaxn & 819^{+10}_{-12} &
1713^{+35}_{-34} & 977^{+4}_{-4} & 3779^{+73}_{-80} \\
\aamidn & 857^{+10}_{-13} & 1952^{+39}_{-50} & 988^{+3}_{-4} & 4418^{+71}_{-60}
&&& \bbmidn & 826^{+12}_{-13} & 1745^{+35}_{-45} & 981^{+3}_{-5} &
3956^{+56}_{-67} &&& \ccmidn & 130^{+6}_{-9} & 139^{+9}_{-11} & 160^{+12}_{-10}
& 175^{+13}_{-12} \\
\aaminn & 433^{+16}_{-16} & 568^{+26}_{-24} & 400^{+19}_{-13} & 514^{+25}_{-21}
&&& \bbminn & 853^{+10}_{-13} & 1907^{+38}_{-40} & 760^{+9}_{-14} &
1423^{+28}_{-53} &&& \ccminn & 42^{+7}_{-6} & 43^{+7}_{-7} & 21^{+6}_{-3} &
21^{+6}_{-3} \\
\enddata
\tablecomments{From left to right, the three big columns correspond to models
in which the MBH--host galaxy scaling relations are constrained by the GWB
``observations'' with (left) and without (middle) redshift evolution,  and the
empirically determined scaling relations (right). From top to bottom, we adopt
the $95\%$ upper limit skymap taken from \citet{Mingarelli17}, and those scaled
by improving the \citet{Mingarelli17} sensitivity skymap by a factor of $2$,
$4$, $8$, and $16$, respectively. $\klocsky$ and $\klocsrc$ represent the number
of skies containing at least one detectable source and the total number of
detectable sources  in $1,000$ realizations of the local BBH population (see
Sec.~\ref{sec:methods:individuals:local}). $\kglbsky$ and $\kglbsrc$ represent
the same quantities, but for the global BBH population (see
Sec.~\ref{sec:methods:individuals:global}). In the table,  the median value and
$16\%$--$84\%$ quantiles of each quantity are listed.   See the details of the
models and the results in Sections~\ref{sec:misc} and
\ref{sec:results:individuals}.}
\end{deluxetable*}
\end{longrotatetable}
\centerwidetable
\begin{longrotatetable}
\begin{deluxetable*}{lRRRRcclRRRRcclRRRR}
%
\tabletypesize{\tiny}
\tablecaption{
Detection prospects of individual BBHs by future PTAs \label{tab:tng}}
\tablehead{
\multicolumn{5}{c}{$z$-independent} &&&
\multicolumn{5}{c}{$z$-dependent} &&&
\multicolumn{5}{c}{Empirical relations} \\
\cline{1-5} \cline{8-12} \cline{15-19}
\colhead{Model} &
\dcolhead{\klocsky} & \dcolhead{\klocsrc} &
\dcolhead{\kglbsky} & \dcolhead{\kglbsrc} &&&
\colhead{Model} &
\dcolhead{\klocsky} & \dcolhead{\klocsrc} &
\dcolhead{\kglbsky} & \dcolhead{\kglbsrc} &&&
\colhead{Model} &
\dcolhead{\klocsky} & \dcolhead{\klocsrc} &
\dcolhead{\kglbsky} & \dcolhead{\kglbsrc}}
\startdata
\multicolumn{19}{c}{{Sensitivity curve of conservative-CPTA}}\\ \hline
\aamaxd & 965^{+6}_{-6} & 3391^{+40}_{-58} & 1000^{+0}_{-0} & 33341^{+179}_{-184} &&&
\bbmaxd & 983^{+3}_{-5} & 4108^{+58}_{-67} & 1000^{+0}_{-0} & 20200^{+136}_{-128} &&&
\ccmaxd & 481^{+12}_{-17} & 651^{+21}_{-21} & 668^{+15}_{-16} & 1106^{+29}_{-33} \\ 
\aamidd & 750^{+13}_{-17} & 1389^{+37}_{-47} & 982^{+3}_{-5} & 3948^{+65}_{-70} &&&
\bbmidd & 671^{+17}_{-16} & 1117^{+36}_{-36} & 930^{+7}_{-8} & 2675^{+49}_{-64} &&&
\ccmidd & 66^{+9}_{-5} & 69^{+8}_{-7} & 71^{+8}_{-7} & 74^{+8}_{-8} \\
\aamind & 349^{+16}_{-14} & 429^{+22}_{-21} & 385^{+12}_{-17} & 487^{+17}_{-23} &&&
\bbmind & 401^{+11}_{-17} & 513^{+16}_{-24} & 494^{+15}_{-16} & 683^{+26}_{-27} &&&
\ccmind & 23^{+5}_{-4} & 23^{+5}_{-4} & 9^{+3}_{-2} & 9^{+3}_{-2} \\
\aamaxn & 974^{+3}_{-6} & 3574^{+57}_{-43} & 1000^{+0}_{-0} & 27627^{+154}_{-197} &&&
\bbmaxn & 726^{+14}_{-16} & 1293^{+37}_{-34} & 1000^{+0}_{-0} & 14981^{+121}_{-149} &&&
\ccmaxn & 710^{+13}_{-15} & 1230^{+39}_{-26} & 943^{+7}_{-8} & 2875^{+53}_{-72} \\
\aamidn & 754^{+16}_{-11} & 1415^{+24}_{-46} & 967^{+5}_{-7} & 3382^{+49}_{-57} &&&
\bbmidn & 722^{+13}_{-19} & 1268^{+41}_{-32} & 950^{+7}_{-6} & 3017^{+43}_{-60} &&&
\ccmidn & 102^{+11}_{-10} & 108^{+12}_{-11} & 127^{+12}_{-12} & 135^{+12}_{-10} \\
\aaminn & 351^{+14}_{-17} & 426^{+27}_{-20} & 329^{+15}_{-15} & 399^{+19}_{-20} &&&
\bbminn & 746^{+12}_{-11} & 1375^{+31}_{-43} & 661^{+10}_{-21} & 1067^{+34}_{-29} &&&
\ccminn & 35^{+5}_{-5} & 35^{+7}_{-5} & 17^{+4}_{-3} & 17^{+5}_{-3} \\  \hline
\multicolumn{19}{c}{{Sensitivity curve of conservative-SKAPTA}}\\ \hline
\aamaxd & 997^{+2}_{-2} & 5786^{+75}_{-75} & 1000^{+0}_{-0} & 72261^{+265}_{-354} &&&
\bbmaxd & 999^{+1}_{-0} & 7316^{+91}_{-102} & 1000^{+0}_{-0} & 45969^{+273}_{-198} &&&
\ccmaxd & 703^{+17}_{-12} & 1220^{+27}_{-38} & 921^{+8}_{-7} & 2544^{+42}_{-61} \\ 
\aamidd & 920^{+10}_{-9} & 2551^{+57}_{-56} & 1000^{+0}_{-0} & 9085^{+100}_{-112} &&&
\bbmidd & 882^{+9}_{-12} & 2142^{+30}_{-75} & 999^{+1}_{-2} & 6460^{+59}_{-95} &&&
\ccmidd & 138^{+11}_{-10} & 150^{+12}_{-14} & 159^{+11}_{-15} & 173^{+12}_{-16} \\
\aamind & 578^{+12}_{-16} & 860^{+26}_{-27} & 685^{+13}_{-21} & 1152^{+33}_{-36} &&&
\bbmind & 629^{+16}_{-17} & 988^{+31}_{-31} & 787^{+14}_{-13} & 1556^{+32}_{-41} &&&
\ccmind & 46^{+7}_{-7} & 48^{+7}_{-8} & 21^{+4}_{-4} & 21^{+5}_{-4} \\
\aamaxn & 998^{+1}_{-2} & 5952^{+61}_{-79} & 1000^{+0}_{-0} & 59788^{+178}_{-302} &&&
\bbmaxn & 905^{+8}_{-9} & 2358^{+46}_{-45} & 1000^{+0}_{-0} & 35772^{+171}_{-212} &&&
\ccmaxn & 888^{+11}_{-11} & 2181^{+52}_{-41} & 998^{+1}_{-1} & 6471^{+90}_{-94} \\
\aamidn & 920^{+11}_{-5} & 2564^{+38}_{-54} & 1000^{+0}_{-1} & 7834^{+98}_{-105} &&&
\bbmidn & 901^{+9}_{-9} & 2304^{+47}_{-45} & 999^{+1}_{-1} & 6821^{+78}_{-91} &&&
\ccmidn & 202^{+11}_{-15} & 224^{+15}_{-14} & 273^{+14}_{-10} & 318^{+21}_{-12} \\
\aaminn & 575^{+12}_{-15} & 847^{+35}_{-28} & 619^{+15}_{-16} & 964^{+28}_{-26} &&&
\bbminn & 920^{+8}_{-7} & 2518^{+55}_{-41} & 904^{+8}_{-13} & 2317^{+56}_{-47} &&&
\ccminn & 70^{+8}_{-8} & 73^{+9}_{-10} & 41^{+5}_{-7} & 42^{+5}_{-8} \\ \hline \hline
\multicolumn{5}{c}{$z$-independent} &&&
\multicolumn{5}{c}{$z$-dependent} &&&
\multicolumn{5}{c}{Empirical relations} \\ \cline{1-5} \cline{8-12} \cline{15-19}
\colhead{Model} &
\dcolhead{\nlocskymean} & \dcolhead{\nlocsrcmean} &
\dcolhead{\nglbskymean} & \dcolhead{\nglbsrcmean} &&&
\colhead{Model} &
\dcolhead{\nlocskymean} & \dcolhead{\nlocsrcmean} &
\dcolhead{\nglbskymean} & \dcolhead{\nglbsrcmean} &&&
\colhead{Model} &
\dcolhead{\nlocskymean} & \dcolhead{\nlocsrcmean} &
\dcolhead{\nglbskymean} & \dcolhead{\nglbsrcmean}\\ \hline
\multicolumn{19}{c}{{Sensitivity curve of optimistic-CPTA}}\\ \hline
\aamaxd & 1.00 & 2020   & 1.00 & 711226 &&&
\bbmaxd & 1.00 & 4687   & 1.00 & 658531 &&&
\ccmaxd & 1.00 & 1581   & 1.00 & 65164  \\ 
\aamidd & 1.00 & 2577   & 1.00 & 212149 &&&
\bbmidd & 1.00 & 2868   & 1.00 & 204405 &&&
\ccmidd & 1.00 & 696    & 1.00 & 8187   \\
\aamind & 1.00 & 1922   & 1.00 & 51281  &&&
\bbmind & 1.00 & 1769   & 1.00 & 41896  &&&
\ccmind & 1.00 & 182    & 1.00 & 798    \\
\aamaxn & 1.00 & 2089   & 1.00 & 655622 &&&
\bbmaxn & 1.00 & 2127   & 1.00 & 469602 &&&
\ccmaxn & 1.00 & 1481   & 1.00 & 111955 \\
\aamidn & 1.00 & 2095   & 1.00 & 182566 &&&
\bbmidn & 1.00 & 2054   & 1.00 & 141318 &&&
\ccmidn & 1.00 & 762    & 1.00 & 13065  \\
\aaminn & 1.00 & 1656   & 1.00 & 43657  &&&
\bbminn & 1.00 & 2182   & 1.00 & 33098  &&&
\ccminn & 1.00 & 227    & 1.00 & 1372   \\ \hline
\multicolumn{19}{c}{{Sensitivity curve of optimistic-SKAPTA}}\\ \hline
\aamaxd & 1.00 & 3921   & 1.00 & 4213492 &&&
\bbmaxd & 1.00 & 11373  & 1.00 & 4213838 &&&
\ccmaxd & 1.00 & 4793   & 1.00 & 554767  \\ 
\aamidd & 1.00 & 6757   & 1.00 & 1671546 &&&
\bbmidd & 1.00 & 7985   & 1.00 & 1683315 &&&
\ccmidd & 1.00 & 2643   & 1.00 & 92162   \\
\aamind & 1.00 & 6062   & 1.00 & 524924  &&&
\bbmind & 1.00 & 5395   & 1.00 & 396741  &&&
\ccmind & 1.00 & 739    & 1.00 & 7703    \\
\aamaxn & 1.00 & 4683   & 1.00 & 4361618 &&&
\bbmaxn & 1.00 & 5825   & 1.00 & 3299998 &&&
\ccmaxn & 1.00 & 4162   & 1.00 & 884334  \\
\aamidn & 1.00 & 5670   & 1.00 & 1537738 &&&
\bbmidn & 1.00 & 5610   & 1.00 & 1209601 &&&
\ccmidn & 1.00 & 2732   & 1.00 & 137059  \\
\aaminn & 1.00 & 5121   & 1.00 & 461490  &&&
\bbminn & 1.00 & 5944   & 1.00 & 263286  &&&
\ccminn & 1.00 & 856    & 1.00 & 12509   \\
\enddata
\tablecomments{ {\scriptsize
The models in this table have similar meanings as those in
Table~\ref{tab:tcg}, excep that the sensitivity curves of the conservative-CPTA,
conservative-SKAPTA, optimistic-CPTA, and optimistic-SKAPTA are adopted,
respectively (see Tab.~\ref{tab:PTAs}).  
For the two conservative PTA configurations, $\klocsky$ and $\klocsrc$
represent the number of skies containing at least one detectable source and the
total number of detectable sources in 1,000 realizations of the local BBH
population (see Sec.~\ref{sec:methods:individuals:local}); while $\kglbsky$ and
$\kglbsrc$ represent the same quantities, but for the global BBH population (see
Sec.~\ref{sec:methods:individuals:global}).
For the two optimistic PTA configurations,  the corresponding quantities denoted
by $\langle\calN\rangle$ (instead of $\calK$) show the average results for one
single realization (instead of the results of the total $1,000$ realizations) of
the BBH populations.}}
\end{deluxetable*}
\end{longrotatetable}

\end{document}